\documentclass[a4paper,11pt]{article}
\pdfoutput=1 
\usepackage{jcappub}
\usepackage{mathtools}
\usepackage{natbib}
\usepackage{multirow}
\usepackage{amsmath}
\usepackage[amssymb]{SIunits}
\usepackage{empheq}
\usepackage{booktabs}

\usepackage{color}

\usepackage{amsfonts}
\usepackage{dsfont} 
\usepackage{graphicx}
\usepackage{amsthm}
\usepackage{calc}
\usepackage{units}
\usepackage{cancel}
\allowdisplaybreaks
\usepackage{lmodern}

\usepackage{csquotes}
\usepackage{varioref}

\usepackage{listings}

\usepackage{booktabs}

\usepackage[svgnames, hyperref]{xcolor}
\newcommand{\integral}[4]{\int\limits_{#2}^{#3} \! #1 \; #4} 
\newcommand{\diff}[1]{\mathrm{d}#1} 

\newcommand{\TM}{T_{\textrm{M}}}
\newcommand{\TCMB}{T_{\textrm{CMB}}}
\newcommand{\sigmav}{\langle\sigma_{\textrm{ann}} v\rangle}
\newcommand{\mDM}{m_{\textrm{DM}}}

\newcommand{\CLASS}{\texttt{CLASS}}
\newcommand{\ExoCLASS}{\texttt{ExoCLASS}}
\newcommand{\DarkAges}{\texttt{DarkAges}}

\usepackage[T1]{fontenc} 

\title{ Exotic energy injection with \ExoCLASS: Application to the Higgs portal model and evaporating black holes}

\author[a]{Patrick St\"ocker,}
\author[a]{Michael Kr\"amer,}
\author[a]{Julien Lesgourgues,}
\author[b]{Vivian Poulin}

\affiliation[a]{Institute for Theoretical Particle Physics and Cosmology (TTK), \\ RWTH Aachen University, D-52056 Aachen, Germany.}
\affiliation[b]{Department of Physics and Astronomy, Johns Hopkins University, Baltimore, MD 21218, USA.}
\emailAdd{stoecker@physik.rwth-aachen.de}
\emailAdd{mkraemer@physik.rwth-aachen.de}
\emailAdd{vpoulin@jhu.edu}
\emailAdd{lesgourg@physik.rwth-aachen.de}

\abstract{We devise a new user-friendly tool interfaced with the Boltzmann code \CLASS~to deal with any kind of exotic electromagnetic energy injection in the universe and its impact on anisotropies of the Cosmic Microwave Background. It makes use of the results from standard electromagnetic cascade calculations develop in the context of WIMP annihilation, generalized to incorporate any injection history. We first validate it on a specific WIMP scenario, the Higgs Portal model, confirming that the standard effective on-the-spot treatment is accurate enough. We then analyze  the more involved example of evaporating Primordial Black Holes (PBHs) with masses in the range $[3\times10^{13},5\times10^{16}]$g, for which the standard approximations break down. We derive robust CMB bounds on the relic density of evaporating PBHs, ruling out the possibility for PBHs with a monochromatic distribution of masses in the range $[3\times10^{13},2.5\times10^{16}]$g to represent all of the Dark Matter in our Universe. Remarkably, we confirm with an accurate study that the CMB bounds are several orders of magnitude stronger than those from the galactic gamma-ray background in the range $[3\times10^{13},3\times10^{14}]$g. A future CMB experiment like CORE+, or an experiment attempting at measuring the 21 cm signal from the Dark Ages could greatly improve the sensitivity to these models.}

\begin{document}
\hfill{\small TTK-XX}\\
\maketitle
\flushbottom

\section{Introduction}
There are nowadays a wealth of observational evidence on a variety of scales in favor of the existence of 85\% of matter  in our universe  in the form of a cold, non-interacting component called Dark Matter (DM) \cite{Bertone:2004pz}. Despite decades of experimental and theoretical efforts at unveiling this mystery, and many potential hints of its detection along the years, we are still lacking a clear non-gravitational identification of the DM, which would potentially hold a lot of information about its nature. Nowadays, the only property commonly accepted is its relic density today $\omega_{\rm cdm}\equiv \rho_{\rm cdm}h^2/\rho_{\rm crit} = 0.1205 \pm 0.0014$ \cite{Aghanim:2016yuo}.
 In the past, searches have focused on the {\em WIMP paradigm}, motivated by the theoretically appealing following observation: a single new particle whose mass and coupling to the standard model (SM) are close to those of the weak sector, will experience a standard chemical decoupling in the early universe (the {\em freeze-out process}) naturally leading to such a relic density today.
This has led to the development of three main channels of detection known as {\em direct detection} (where one tries to measure the recoil of a nucleon, eventually electron, being hit by a DM particle), {\em indirect detection} (where one tries to measure the anti-particles produced by residual annihilations of DM) and {\em production at colliders} (where one tries to measure the DM ``missing energy'' when reconstructing the invariant mass of particles produced during the collision).  
Remarkably, cosmological probes and in particular the Cosmic Microwave Background (CMB) can be used to perform both direct and indirect detection of DM. Indeed, non-gravitational interactions between the DM and the SM may affect the growth of perturbations, leading to clear signatures on the CMB temperature and polarization power spectra \cite{Dvorkin:2013cea,Wilkinson:2013kia,Wilkinson:2014ksa,Gluscevic:2017ywp}. On the other-hand, any energy injection from DM annihilation (or decay) affects the evolution of the free electron fraction $x_e \equiv n_e/n_H$ and of the intergalactic medium temperature $\TM$. For large enough energy injection rates, this mechanism could also have a strong impact on the CMB power spectra \cite{Padman05,Hooper09,Cirelli09,Huetsi:2009ex,Slatyer09,Natarajan08,Natarajan09,Natarajan10,Valdes:2009cq,Evoli:2012zz,Galli13,Finkbeiner11,Hutsi:2011vx,Slatyer12,Giesen,Slatyer13,Slatyer15-1,Lopez-Honorez:2013lcm,Poulin:2015pna,Liu:2016cnk}.

It has been shown in the past that the phenomenology associated to exotic electromagnetic energy injection is extremely rich (see in particular Ref.~\cite{Finkbeiner11,Slatyer:2016qyl,Poulin2016} for a detailed analysis). Indeed, from the study of the CMB power spectra, it is possible to learn not only on the {\em amount} of electromagnetic (e.m.) energy injection but also on its {\em time dependence}, and thus to potentially distinguish between different scenarios leading to exotic e.m. energy injection.

In this paper, we  develop and release a tool calculating the impact on the CMB of virtually any form of exotic electromagnetic energy injection. 
It basically ``fills the gap'' between existing public codes calculating, on the one hand, the development of the e.m. cascade in the plasma and its impact on the free-electron fraction, and on the other hand, the CMB power spectra.  It makes use of the results from Refs.~\cite{Slatyer09,Slatyer15-2}, developed in the context of WIMP annihilation, and it broadens their range of application to an arbitrary injection history, or in other words, to any particle injection spectrum and injection rate. With such a tool, it is possible to study for instance DM annihilation (including the impact of halo formation at low-$z$), DM decay (even when only a fraction of the total DM is unstable), or even Primordial Black Holes (PBHs) evaporation or accretion. 
We discuss in details two particular examples, which had only been studied using an effective parametrisation or some crude approximations: first, the scalar Higgs portal DM model, and second, low-mass evaporating PBHs with $M\lesssim10^{17}$g. 
The first example allows us to validate our tool on a slightly non-trivial scenario and can be regarded as a ``proof-of-principle''. It confirms previous results showing that in the context DM annihilation searches, one does not need to follow accurately the full redshift dependence of the energy deposition function \cite{Finkbeiner11}. However, the second example is less trivial, given that in the case of an energy injection rate scaling like $\sim (1+z)^3$, any miscalculation of the energy deposited in the plasma leads potentially to large inaccuracies in CMB predictions, in particular when the peak of the energy injection is around recombination \cite{Poulin2016}. Moreover,  when PBHs evaporate, both the {\it production rate} and the {\it spectrum} of injected particles change with time, a complication which has always been neglected until now when calculating the CMB spectra.

The paper is structured as follows. In section \ref{sec:2}, we first recall basic principles on e.m. energy injection and on its impact on the free-electron fraction and IGM temperature. We then detail the content and the use of the new \ExoCLASS~branch of the Boltzmann code \CLASS\footnote{https://www.class-code.net} \cite{Class2011_b}, which incorporates a python module named \DarkAges~that calculates the energy deposited in the plasma. We also describe how one can easily implement any new injection history not yet coded in the \DarkAges~module.
In section \ref{sec:HiggsPortal}, we study the case of the scalar Higgs portal DM model. We compute the most up-to-date cosmological constraints on the DM mass and coupling to the standard model in this scenario, and we validate our tool by comparing our accurate approach to the standard approximation (the so-called ``effective on-the-spot'' treatment). In section \ref{sec:PBH}, we study the more involved case of evaporating PBHs. We derive CMB bounds on the fraction of DM in the form of PBHs as a function of the PHB mass (for a population of monochromatic PBHs), and we compare these bounds to constraints coming from Extragalactic Gamma-ray Background (EGB) studies. Finally we report our conclusions in section \ref{sec:conclu}.

\section{The \ExoCLASS~branch of the \CLASS{} code}
\label{sec:2}
\ExoCLASS{}  is a public branch of the \CLASS{} code aimed at computing the CMB power spectra in the presence of an electromagnetic cascade (e.m. cascade) caused by any source of injection of electromagnetic energy. 
\ExoCLASS{} implements several new features with respect to the main branch:
\begin{itemize}
\item first, the user can specify several injection histories in the form of different expressions for the function
$$\frac{dE}{dVdt}\bigg |_{\textrm{inj}}(z)$$ that will be defined in equation (\ref{eq:dep_inj}). In section 2.2 we will review different possibilities and their theoretical motivations. The main branch of \CLASS{} only incorportates a simplified way to treat energy injection in the specific case of $s$-wave DM annihilation.
\item 
a new package written in python, called \DarkAges, which role is to compute the efficiency function in different channels $f_c(z)$ that will be defined later in section 2.2, based on the method developed in~\cite{Slatyer09,Slatyer13,Slatyer15-2} and on the integral of equation~(\ref{eq:fz_anni}). This module must be interfaced at least with the public tool kit of Ref.~\cite{Slatyer15-2}\footnote{\url{http://nebel.rc.fas.harvard.edu/epsilon/}}. It also needs some input for the spectrum of electrons, positrons, photons and other annihilation or decay products, that can be either computed from first principles (in this work we used the code \texttt{PYTHIA\footnote{\url{http://home.thep.lu.se/~torbjorn/pythia8/pythia8219.tgz}} v8.219} \cite{Sjostrand:2014zea}), or taken as tabulated templates e.g. from the \texttt{PPPC4DMID}\footnote{\url{http://www.marcocirelli.net/PPPC4DMID.html}} \cite{Cirelli2010,Ciafaloni:2010ti}. \DarkAges{} can be used in two ways: as a stand-alone code, or as an internal \CLASS{} module. We describe its structure and its basic usage in section 2.3.
\item finally, some modifications to the {\tt thermodynamics} module of \CLASS{}, that are necessary for calling \DarkAges, as well as new interfaces with the latest available versions of the recombination codes \texttt{HyRec}  \cite{AliHaimoud2011}\footnote{\url{http://cosmo.nyu.edu/yacine/hyrec/hyrec.html}} and  \texttt{CosmoRec}  \cite{Chluba2010} \footnote{\url{http://www.cita.utoronto.ca/~jchluba/Science_Jens/Recombination/CosmoRec.html}}.
\end{itemize}
A few days after the submission of this paper, the \ExoCLASS{} branch will be available for cloning or download from the \CLASS{} GitHub repository\footnote{\url{https://github.com/lesgourg/class_public}}. In the future, some of these features
(in particular, the interface with the latest \texttt{HyRec} and \texttt{CosmoRec}) will be propagated to the main {\tt master} branch.

\subsection{Theoretical basics}

The main impact of exotic energy injection on the CMB is related to a modification of the free electron fraction evolution. In particular, the modification of the {\em Thompson optical depth} $\tau(z)$ and  {\em visibility function} $g(z)$ around and below  $z\sim1000$ will lead to very distinct features on the CMB power spectra, which just arise from modifications to the recombination  equations.

We refer to electromagnetic energy injection as a generic word for the injection of photons and electromagnetically charged particles from the standard model (SM) following processes such as annihilations or decays. On the cosmological timescales of interest, all unstable particles will decay producing a primary spectrum of stable particles whose interaction with the cosmological plasma need to be accurately followed. Like in most of the literature, we restrict the analysis to the impact of injected positrons, electrons  and photons. Indeed, neutrinos are basically invisible to the medium and simply carry away part of the energy. On the other hand, protons and antiprotons have been checked to loosen the bounds by about 10\%, \cite{Weniger:2013hja}. Thus neglecting them only leads to slightly too conservative bounds, while permitting a significant reduction of the computing time. 

Typically, the injected primary particles initiate an  e.m. cascade by interacting with thermal photons, producing an increase in the number of non-thermal particles at the expense of a decrease in their average energy. When these extra particles cool down to energies of the order of a keV, they start interacting strongly with atoms of hydrogen (and sub-dominantly of helium \cite{Galli13}). To account for the ionization, excitation and heating of these atoms, we have to modify the equations governing the evolution of the fraction of free electrons, $x_e\equiv n_e/n_H$, taking into account both direct ionization and collisional excitation followed by photoionization by a CMB photon. At the same time, we must add to the equation for the evolution of the intergalactic medium (IGM) $T_M$ a term accounting for the associated heating, which has a feedback on the evolution of $x_e$. Finally, at some point, the energy of the extra particles drops below the Lyman-$\alpha$ transition energy (10.2 eV). Then these particles are no longer able to interact with atoms and can be considered as ``lost''. The
{\it three-level atom} approximation gives a good overall description of the processes at play, and can be fudged to achieve sub-percent accuracy \cite{RubinoMartin:2009ry}. In this approximation, the evolution equations of the free electron fraction $x_e$ and IGM temperature $\TM$ is governed by:

 \begin{eqnarray}\label{eq:x_e&T_M}
\frac{dx_{e}(z)}{dz}=\frac{1}{(1+z)H(z)}(R(z)-I(z)-I_X(z))~,\nonumber\\
\frac{d\TM}{dz}  =  \frac{1}{1+z}\bigg[2\TM+\gamma(\TM-\TCMB)\bigg]+K_h~.
\end{eqnarray}
where the $R$ and $I$ terms are the standard recombination and ionization rates given by
\begin{equation}\label{eq:RandI}
R(z) = C\bigg[\alpha_{H} x_e^2 n_H\bigg], \qquad I(z)  = C\bigg[ \beta_{H}(1-x_e)e^{-\frac{h\nu_\alpha}{k_b\TM}}\bigg].
\end{equation}
The effective ionization rate $I_X$ can be decomposed as
 $I_X(z) = I_{Xi}(z)+I_{X\alpha}(z)$, where $I_{Xi}$  is the rate of direct ionization and $I_{X{\alpha}}$  that of excitation+ionization:
\begin{equation}\label{eq:IonizationRate}
I_{Xi} = -\frac{1}{n_H(z)E_i}\frac{dE}{dVdt}\bigg |_{\textrm{dep}, i}~, \qquad I_{X{\alpha}} =  -\frac{(1-C)}{n_H(z)E_{\alpha}}\frac{dE}{dVdt}\bigg |_{\textrm{dep}, \alpha}~,
\end{equation}
while $E_i$ and $E_\alpha$ are respectively the average ionization energy per baryon, and the Lyman-$\alpha$ energy.
Finally, the rate $K_h$ at which the plasma is heated by DM decay or annihilation is defined as:
\begin{equation}\label{eq:Kh}
K_h =-\frac{2}{H(z)(1+z)3k_b n_H(z)(1+f_{He}+x_e)}\frac{dE}{dVdt}\bigg |_{\textrm{dep}, h}\quad.
\end{equation}
We refer e.g. to the appendix of Ref.~\cite{Poulin2016} for further definitions and more details on each of these coefficients. In \CLASS, it is possible to use a fudged version of Recfast \cite{Seager1999_a,RubinoMartin:2009ry} or \texttt{HyRec} code \cite{AliHaimoud2011} to solve these recombination equations. The \ExoCLASS{} branch proposes as a third possibility the use of \texttt{CosmoRec}  \cite{Chluba2010}.

All these processes have an impact on the CMB anisotropy angular power spectra through an enhancement of Thomson interactions between CMB photons and free electrons. In particular, the modified free electron fraction affects two very important thermodynamics quantities entering in the line-of-sight solution of the CMB photon Boltzmann hierarchy \cite{Seljak:1996is}, namely the {\em Thompson optical depth} $\tau_{\rm depth}(z)$ and the {\em visibility function} $g(z)$,
\begin{eqnarray}\label{eq:opt_depth_vis_function}
\tau_{\rm depth}(z) \equiv  \int^{z}_{0} n_H(z)x_e(z) \sigma_T\frac{dt}{dz'} dz'~\,,\qquad
g(z) \equiv e^{-\tau}\frac{d\tau_{\rm depth}}{dz}\,.
\end{eqnarray}

In order to compute the three rates defined in equations (\ref{eq:IonizationRate}, \ref{eq:Kh}), we need to know the rate of energy density \emph{deposition} in the plasma at redshift $z$, $\frac{dE}{dVdt} \big |_{\textrm{dep}}(z)$, and how it is splitted between ionization, excitation of the  Lyman-$\alpha$ transition, and heating of the IGM. It can be linked to the rate of energy injection per unit volume through three dimensionless functions $f_c(z)$:
\begin{equation}\label{eq:dep_inj}
\frac{dE}{dVdt}\bigg |_{\textrm{dep},c}\!\!\!\!\!\!\! (z) =f_c(z)\frac{dE}{dVdt}\bigg |_{\textrm{inj}} \!\!\! (z)~,
\end{equation}
where the index $c$ runs over the three channels called ($i$, $\alpha$, $h$) in equations (\ref{eq:IonizationRate}, \ref{eq:Kh}). The sum of the three components of $f_c(z)$ should always be smaller than one, to account for the fraction of injected energy ending up in very low energy photons ($\lesssim$10.2 eV) which are unable to interact.

In the standard formalism, the {\em energy deposition functions} $f_c(z)$ can be obtained by convolving a given primary injection spectrum at redshift $z'$, $\frac{dN}{dE} |^{(\ell)}_\mathrm{inj}(z',E)$, with a set of {\em transfer functions} $T_c^{(\ell)}(z',z,E) $ defined as the fraction of the energy $E$ injected at $z'$ that is deposited at $z$ in a given channel $c$ for a given particle $\ell=\gamma,e^\pm$. 
The main difficulty in solving the problem at hand is that,  in order to correctly compute these transfer functions, it is necessary to follow the evolution of the daughter particle spectra over a very large range of energy and time scales.
In particular, it has been shown in Ref.~\cite{Slatyer09} that for WIMP annihilation, at redshift around and below recombination, the injected energy is usually not absorbed {\em on-the-spot} (at $z\simeq z'$), since at these times cooling processes are unefficient over a Hubble time scale~\cite{Slatyer09}. Technically, this means that the Boltzmann equations describing the evolution of the each particle spectrum cannot be solved in a stationary approximation, and thus require more involved resolution schemes such as Monte Carlo methods (e.g. ~\cite{Valdes:2009cq,Evoli:2012zz,Slatyer09}). Reference~\cite{Poulin2016} extended this analysis to the case of decaying particles with arbitrary lifetimes. Through a detailed comparison of several energy deposition schemes, this work shows that an accurate determination of the energy deposition function is mandatory in order to get reliable predictions for the impact of energy injection on the CMB.

In this work, we use the most up-to-date transfer functions computed in Ref.~\cite{Slatyer15-1,Slatyer15-2}. These functions were derived in the context of WIMP annihilation, but they actually have a wider range of application, as we shall discuss later. We refer the reader interested in understanding the resolution methods and the associated systematic uncertainties to~\cite{Slatyer15-1,Slatyer15-2}. We simply recall that one of the main assumptions in such calculations is that the altered free electron fraction will not significantly back-react onto the energy cascade evolution. One could improve over this approximation by computing some $x_e$-dependent transfer functions. We do not address such complications, given that no sign of exotic energy injection has been detected yet. Moreover, it has been shown in Ref.~\cite{Liu:2016cnk} that neglecting such a back-reaction leads to conservative constraints.

\subsection{Implementation of various injection histories in \ExoCLASS}

We have implemented in \ExoCLASS{} most of the energy injection histories that have been studied so far:
\begin{itemize}
\item {\em DM-annihilation, including the impact of halo formation at low-$z$}: The impact of  DM annihilation onto the CMB  has  been extensively studied in the literature, both in the smooth background   \cite{Padman05,Hooper09,Cirelli09,Huetsi:2009ex,Slatyer09,Natarajan10,Valdes:2009cq,Evoli:2012zz,Galli13,Finkbeiner11,Hutsi:2011vx,Slatyer12,Giesen,Slatyer13,Slatyer15-1} 
and within halos \cite{Hooper09,Huetsi:2009ex,Natarajan08,Natarajan09,Natarajan10,Cirelli09,Giesen,Lopez-Honorez:2013lcm,Poulin:2015pna,Liu:2016cnk}. The rate of energy injection in the smooth background can be written as 
\begin{equation}
\left.\frac{\diff^2 E}{\diff V \diff t} \right\vert_\mathrm{inj,~smooth}  = \kappa \rho^2_c c^2 \Omega_\mathrm{CDM}^2 \left( 1+z \right)^6 \frac{\sigmav}{m_\mathrm{DM}}\,,
\label{eq:injection_anni}
\end{equation}
and involves the critical density today $\rho_c$, the fractional DM density today $\Omega_\mathrm{CDM}$, and the DM mass $m_\mathrm{DM}$ and annihilation cross-section $\sigmav$. If DM is made of self-conjugated particles, such as Majorana fermions, one has $\kappa=1$, while if DM particles and antiparticles differ (as in the case of Dirac fermions) and are equally populated, $\kappa=1/2$.
The main impact of structure formation is to enhance the average squared energy density with respect to the smooth background case,
by an amount usually parametrized through a boost factor $\mathcal{B}(z)$:
\begin{equation}
\langle \rho^2 \rangle (z)=  (1+\mathcal{B}(z)) \,\, \langle {\rho} \rangle^2(z).
\end{equation}
Further details on this function can be found in the appendix C of Ref.~\cite{Poulin:2015pna}. Currently, we have only implemented a simple parametrization based on the Halo model and on the Press-Schechter formalism, but it would be trivial to generalise the expression of $\mathcal{B}(z)$. 
The energy injection rate from annihilation in halos is thus:
\begin{equation}\label{eq:EnergyDepositionSmoothHalos}
\frac{dE}{dVdt}\bigg|_{\textrm{inj,~halos}}=\kappa \rho^2_cc^2\Omega^2_{\textrm{DM}}(1+z)^6\frac{\sigmav}{\mDM}\mathcal{B}(z)~,
\end{equation}
and the total energy injection rate is simply the sum of the two contributions. We  describe how to make use of this injection history in \ExoCLASS{} (and all other histories), the relevant key parameters and available options for the energy deposition in sec.~\ref{sec:ExoClass}. 

More details can also be found in the files \texttt{explanatory.ini} and \texttt{README} of the \ExoCLASS{} branch.

\item {\em DM-decay with arbitrary lifetimes from a fraction of the DM:} The decay of exotic particles and its impact on the CMB has been scrutinized in the recent literature \cite{Chen2003,Kasuya06,Zhang07,Yeung12,Slatyer13,Liu:2016cnk,Oldengott:2016yjc,Slatyer:2016qyl,Poulin2016}. 
The decay of a fraction $\Xi$ of the DM with a constant rate $\Gamma$ and  lifetime $\Gamma^{-1}$ injects some energy density in the plasma at a rate 
\begin{equation}
\left.\frac{\diff^2 E}{\diff V \diff t} \right\vert_\mathrm{inj,~dec.} = \Xi \rho_c c^2 \Omega_\mathrm{CDM} \left( 1+z \right)^3 \Gamma \exp{\left(-\Gamma t\right)}~.
\label{eq:injection_decay}  
\end{equation}
Within \CLASS, the user can pass the fraction of decaying DM $\Xi \equiv \rho_\chi/\rho_{\rm DM}$ and its lifetime $\tau\equiv\Gamma^{-1}$ in units of $s$.  More details on these parameters can be found in the file \texttt{explanatory.ini}. We follow the standard approach and define the {\it deposited} energy with respect to the {\it injected} energy of a long-lived particle:
\begin{equation}
\left.\frac{\diff^2 E}{\diff V \diff t} \right\vert_{\mathrm{dep},c} \!\!\!\!\!\!\!  (z) = f_c(z) \lim_{\Gamma t \ll 1} \left[  \left.\frac{\diff^2 E}{\diff V \diff t} \right\vert_\mathrm{inj,~dec.} \!\!\!\!\!\!\!  (z) \right]~.
\end{equation}
This just means that we absorb the exponential factor within the definition of the $f_c(z)$ function. 

\item {\em Evaporation of light PBHs:}  We will discuss this case in details in sec.~\ref{sec:PBH}  (see also \cite{Poulin2016}).  For a population of monochromatic PBHs accounting for a fraction $ f_\mathrm{PBH} =\rho_{\rm PBH}/\rho_{\rm DM}$ of the DM, each with an initial mass $M_\mathrm{ini}$, an evaporation rate $\frac{\diff M}{\diff t}$ and an electromagnetic branching ratio $f_{\rm e.m.}$, the energy injection rate reads:
\begin{equation}
\left.\frac{\diff^2 E}{\diff V \diff t} \right\vert_\mathrm{inj,~PBH} = \frac{f_\mathrm{PBH} \, f_{\rm e.m.} \, \rho_c \, c^2 \Omega_\mathrm{CDM} \left( 1+z \right)^3}{M_{\rm ini}} \frac{\diff M}{\diff t}\,.
\end{equation}

\item {\em Accretion of matter onto heavy PBHs:} Heavy ($M\gtrsim 1 M_\odot$) PBHs  have received a lot of attention after the aLIGO discovery of binary black hole (BH) mergers of tens of solar masses~\cite{Abbott:2016blz,Abbott:2016nmj,Abbott:2017vtc,Abbott:2017oio,Abbott:2017gyy}. In particular, based on computation of the merging rate of PBHs today, it has been suggested that they could represent a large fraction (if not all) of the DM in our Universe \cite{Bird:2016dcv,Sasaki:2016jop,Clesse:2016vqa,Clesse:2017bsw}. Interestingly, such massive objects accrete matter, which heats up, gets eventually ionized and eventually emits high-energy radiation. This radiation can in turn leave distinct imprints on the CMB power spectra, eventually jeopardizing the success of $\Lambda$CDM, and the CMB can thus put stringent bounds on the fraction of the DM made of PBHs. 
For black holes with relative density $f_{\rm PBH}$ and mass $M$,
the total energy injection rate per unit volume is \cite{Poulin:2017bwe}:
\begin{equation}
\left. \frac{\diff^2 E}{\diff V \diff t} \right\vert_\mathrm{inj,~acc.} =L_{\rm acc}n_{\rm PBH}=L_{\rm acc}f_{\rm PBH}\frac{\rho_c  \Omega_\mathrm{CDM} \left( 1+z \right)^3}{M}\,.
\end{equation}
where $L_{\rm acc}$ is the {\em accretion luminosity}, which is affected by large theoretical uncertainties. Usually, it is written as $L_{\rm acc}=\epsilon \dot{M}$, where 
$ \dot{M}$ is the accretion rate and $\epsilon$ the radiative efficiency factor at which matter is converted to radiation. Moreover, it is necessary to specify an energy spectrum for the radiation, which is typically dominated by Bremsstrahlung emission up to hundreds of keV.
All these quantities can be determined within a given accretion scenario, but the question of which scenario is most relevant is still unclear. We have implemented the spherical accretion scenario of Ref.~\cite{Ricotti:2007au}, as well as its re-evaluated version from Ref.~\cite{Ali-Haimoud:2016mbv}, which corrected several mistakes.
Recently, it has been noted that the hypothesis of spherical accretion might break down already around recombination. Accretion could thus happen in a disk, enhancing greatly the radiative efficiency and in turn the impact onto the CMB power spectra. We have also implemented a disk accretion scenario as advocated in Ref.~\cite{Poulin:2017bwe}, and we refer to this later article for details on these various scenarios.

\end{itemize}

\subsection{The \DarkAges{} module}

The main role of the \DarkAges{}-package~is to compute the energy deposition functions $f_c(z)$ introduced above for any energy injection history, i.e. a spectra of injected particles at a given rate. These functions read in general
\begin{equation}
f_c(z) = \frac{\integral{\diff{\ln{\left(1+z'\right)}}}{z}{\infty}{ \frac{(1+z')^\alpha}{H(z')} \sum_\ell \integral{\diff{E}}{0}{m}{ T_c^{(\ell)} (z',z,E) E \left.\frac{\diff{N} \left(E,t(z)\right)}{\diff{E}\diff{t}}\right\vert^{(\ell)}_\mathrm{inj.}} } }{ \frac{(1+z)^\alpha}{H(z)} \integral{\diff{E}}{0}{m}{ E \left.\frac{\diff{N} \left(E,t(z)\right) }{\diff{E}\diff{t}} \right\vert^\mathrm{tot.}_\mathrm{inj.}} }
\label{eq:fz_anni}
\end{equation}
where $\alpha = 0$ for an energy rate that scales like  $\sim(1+z)^3$ (e.g. for decays) and $\alpha = 3$ if it scales like   $\sim(1+z)^6$ (e.g. for $s$-wave annihilations).
Note that we generalize the original formalism, by using the differential rate of the particle spectrum with time rather than a fixed reference spectrum. In general the time dependence can be factorized, but it is not always the case (e.g. for evaporating PBHs).
The transfer functions have been tabulated for energy injection above 5 keV.  However, for energy injection below that, we can safely extrapolate the transfer function down to $\sim$ 100 eV: It has been shown in Ref.~\cite{Galli13} that the energy repartition functions are to an extremely good approximation independent of the initial particle energy in the range between  $\sim$ 100 eV and a few keV. In fact, this behaviour is at the heart of the ``low energy code''  used by the authors of Ref.~\cite{Slatyer15-2} to compute their transfer functions. Below  $\sim$ 100 eV, the ionization efficiency starts to drop, and we conservatively cut the integral at the numerator at this energy.

The application of equation \eqref{eq:fz_anni} to calculate $f_c(z)$ from a given injected spectral rate and for a given scaling $\sim(1+z)^{\alpha+3}$ is the main purpose of the \DarkAges-module. Apart from that, the module also includes its own interpolation class and routines to calculate the injected spectral rate for the scenarios given in section 2.2. For the sake of modularity and flexibility the module is intentionally written in Python, such that individual functions of the module can be easily used in a different context and to simplify the implementation of custom injection histories.

The \DarkAges-module can be used, independently of \ExoCLASS, as a stand-alone python routine in order to simply compute $f_c(z)$ and print it in a file.  Its usage is explained in details in App.~\ref{sec:DarkAges}, and additional information can be found in the documentation of the module, which is located within the \verb|doc| folder. A basic example can also be found as a Jupyter notebook within the \verb|examples| folder.

Within the \ExoCLASS~branch, there are several ways to compute the impact of electromagnetic energy injection, typically leading to different level of accuracy and runtime. We detail the use of the \ExoCLASS~branch, the various new options available, and how to call the \DarkAges~module within the \texttt{.ini}-file in App.~\ref{sec:ExoClass}.
Moreover, the \ExoCLASS~branch contains additional reionization parametrization, including an asymmetric reionisation as described in Ref.~\cite{DiValentino:2016foa,Adam:2016hgk} and a semi-analytical resolution based on the model of star formation from  Ref.~\cite{Robertson:2015uda} (introduced in Ref.~\cite{Poulin2015} in the context of CMB studies). More details can be found in the file \texttt{explanatory.ini} within the \CLASS~folder of the \ExoCLASS~branch.
\section{First application: the Higgs portal model}
\label{sec:HiggsPortal}
The scalar Higgs portal model~\cite{Silveira:1985rk,McDonald:1993ex,Burgess:2000yq} comprises the Standard Model and a real scalar field, $S$, which is a singlet under all SM gauge groups and odd under a $Z_2$ symmetry, $S \to -S$. The scalar is thus stable and a viable dark matter candidate. 
The Lagrangian of the scalar Higgs portal model is given by 
\begin{equation}
\mathcal{L} = \mathcal{L}_\mathrm{SM} - \frac{1}{2} \partial_\mu S \partial^\mu S - \frac{1}{2} \mu_S^2 S^2 - \frac{1}{4} \lambda_S S^4 + \frac{1}{2} \lambda_{H\!S} S^2 H^\dagger H\,. 
\label{eq:HP_Lag_noEWSB} 
\end{equation} 
After electroweak symmetry breaking, the mass and interaction terms of the scalar field become 
\begin{equation}
\mathcal{L} \subset - \frac{1}{2} m_S^2\, S^2 - \frac{1}{4} \lambda_S\, S^4 - \frac{1}{2} \lambda_{H\!S}\, h^2 S^2 - \frac{1}{2} \lambda_{H\!S}\, v h S^2 \,, 
\label{eq:HP_Lag_noEWSB} 
\end{equation} 
where $H = (h + v, 0)/\sqrt{2}$, $v = 246$~GeV. The physical mass of the scalar dark matter particle, $m_S$, is given by $m_S^2 = \mu_S^2 + \lambda_{H\!S} v^2/2$. The quartic interaction $\propto \lambda_S\, S^4$ affects the stability of the electroweak vacuum, but is not relevant for dark matter phenomenology. For our purposes, the model is therefore specified by only two free parameters, i.e.\ the dark matter mass, $m_S$, and the coupling of dark matter to the SM Higgs, $\lambda_{H\!S}$. 

The scalar Higgs portal model is among the simplest UV-complete dark matter models. While the model is arguably too simplistic in its minimal form, equation~\ref{eq:HP_Lag_noEWSB}, a coupling between a gauge singlet dark sector and the Standard Model via a Higgs bilinear term is expected in a wide class of models, as $H^\dagger H$ is the only SM gauge singlet operator of mass dimension two. 

The Higgs portal interaction $\propto \lambda_{H\!S} S^2 H^\dagger H$ provides an efficient dark matter annihilation mechanism. The Higgs portal model can thus describe the dark matter relic density, and it can be probed in indirect dark matter searches. Both the $\gamma$-ray galactic center excess and a potential dark matter signal in cosmic rays antiprotons can be accommodated by the Higgs portal model~\cite{Cuoco:2017rxb}. Furthermore, the Higgs portal interaction leads to invisible Higgs decays, $h \to SS$, and a DM-nucleon interaction through the exchange of a Higgs boson, which can be probed at the LHC and in direct detection experiments, respectively. For a recent discussion of the various constraints see e.g.~\cite{Cuoco:2016jqt,Athron:2017kgt}. 

\begin{figure}[htb]
 \centering
 \includegraphics[width=.65\textwidth]{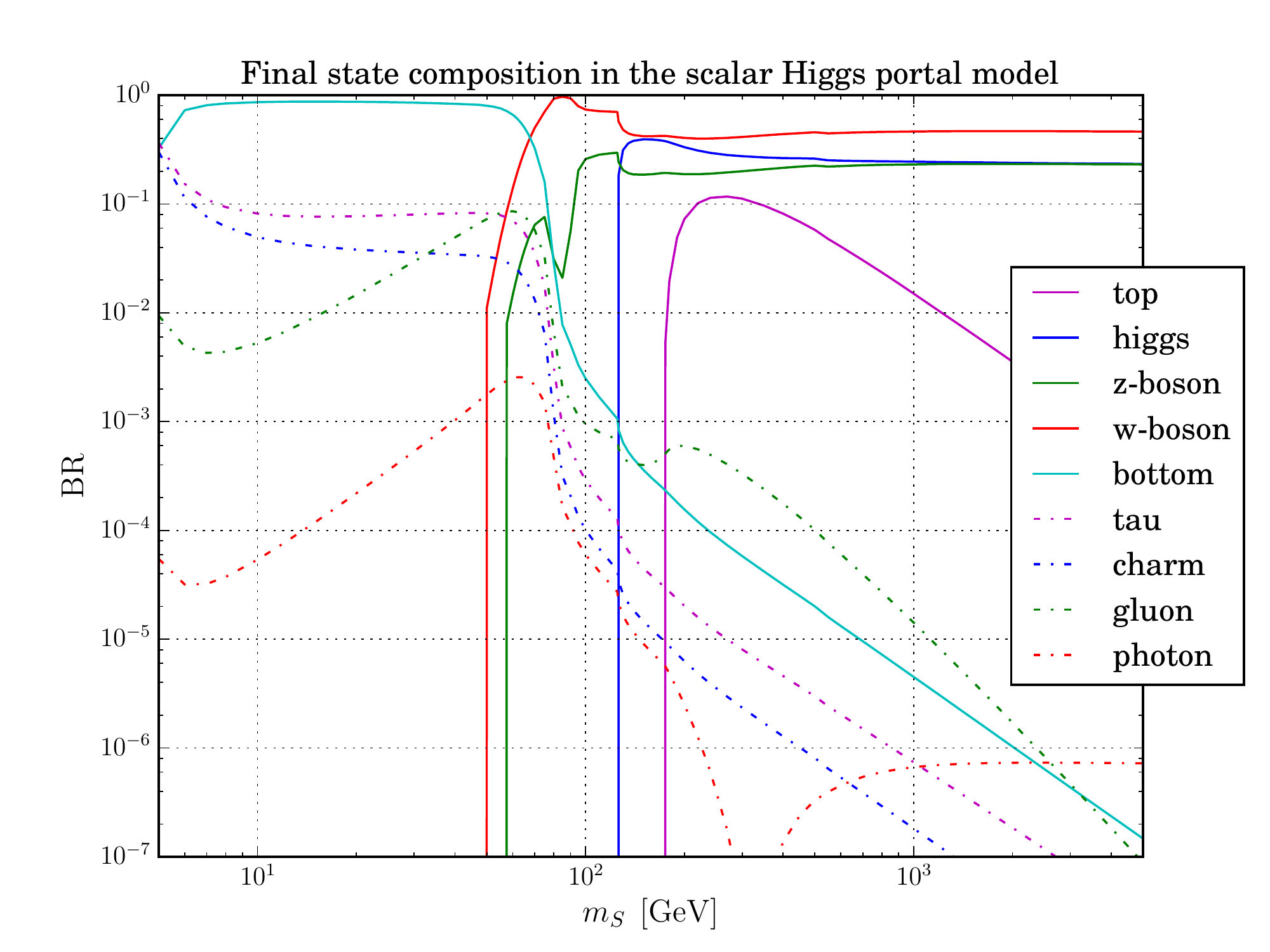}
 \caption{\label{fig:HP_branching}Branching ratios of the various SM final states for the annihilation of scalar dark matter $S$ as a function of its mass $m_S$. The portal coupling $\lambda_{H\!S}$ is fixed to a value that 
 corresponds to the correct relic density $\Omega h^2 = 0.1205$.}
 \end{figure}

Within the Higgs portal model, dark matter annihilation into Standard Model final states proceeds through the exchange of a Higgs boson in the $s$- and $t$-channel, and through the quartic $S^2 h^2$ interaction. Below the Higgs threshold, $m_S < m_h$, only $s$-channel Higgs exchange is relevant, so that the relative weight of the different SM final states is determined by the SM Higgs branching ratios and independent of the portal coupling $\lambda_{H\!S}$. For dark matter masses $m_S \ge m_h$, $hh$ final states contribute and the composition of the SM final state depends on the size of the $\lambda_{H\!S}$. In figure~\ref{fig:HP_branching} we display the relative size of the different SM annihilation channels as a function of the dark matter mass in the range 5~GeV~$< m_S<$~5~TeV. The Higgs portal coupling $\lambda_{H\!S}$ has been chosen such that for a given value of $m_S$ the correct relic density $\Omega h^2 = 0.1205$~\cite{Aghanim:2016yuo} is obtained. (We use micrOMEGAs~\cite{Belanger:2014vza} to calculate the relic density, see also~\cite{Cuoco:2016jqt}.) To compute the annihilation spectrum, we combine the spectra of the various SM final states as provided in~\cite{Cirelli:2010xx}, weighted by their relative strength as displayed in figure~\ref{fig:HP_branching}. We also include the contributions of three-body final states from the annihilation into off-shell gauge bosons, see~\cite{Cuoco:2016jqt}. 

\begin{figure}[htb]
\centering
\includegraphics[width=.45\textwidth]{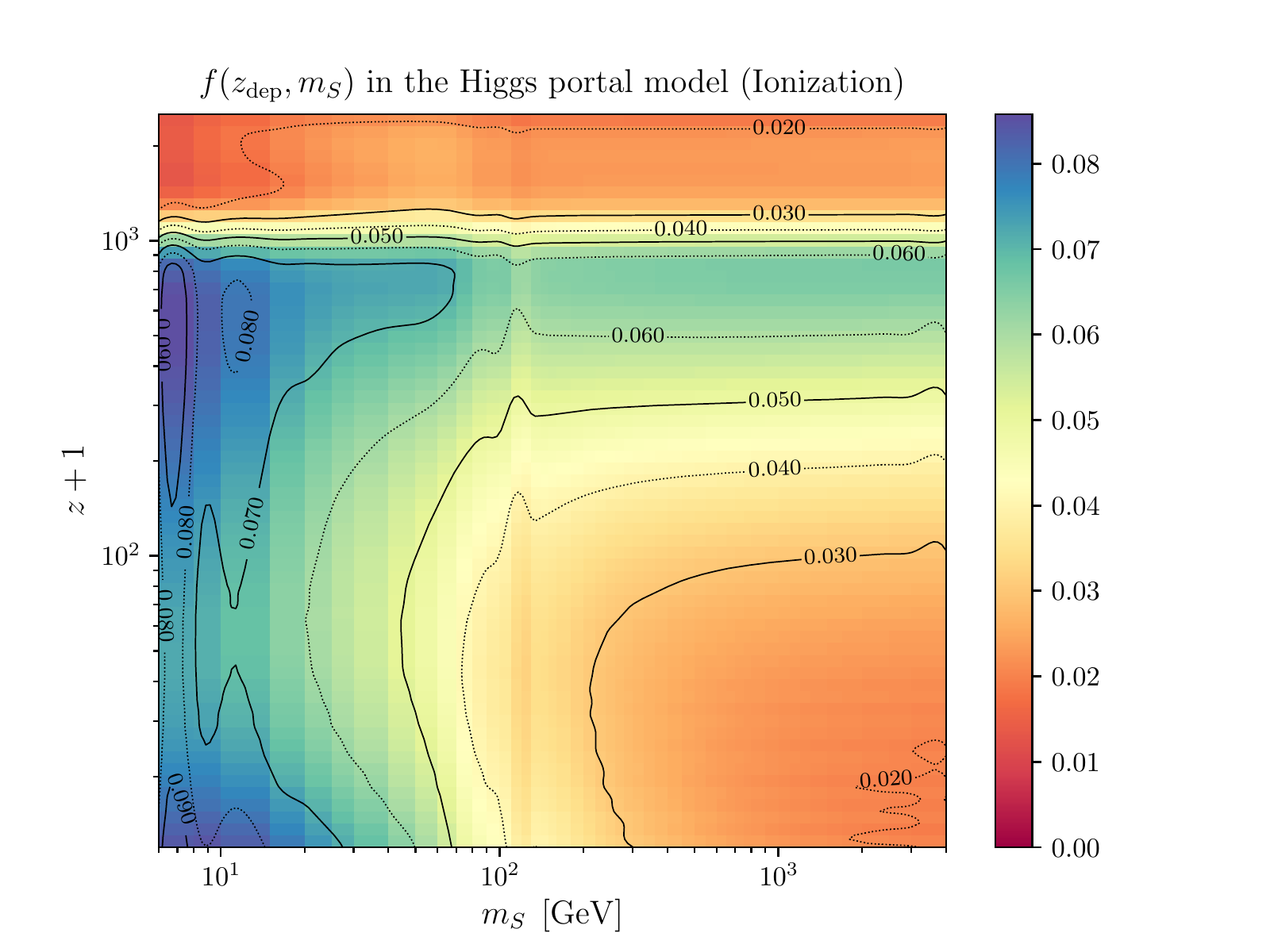}
\includegraphics[width=.45\textwidth]{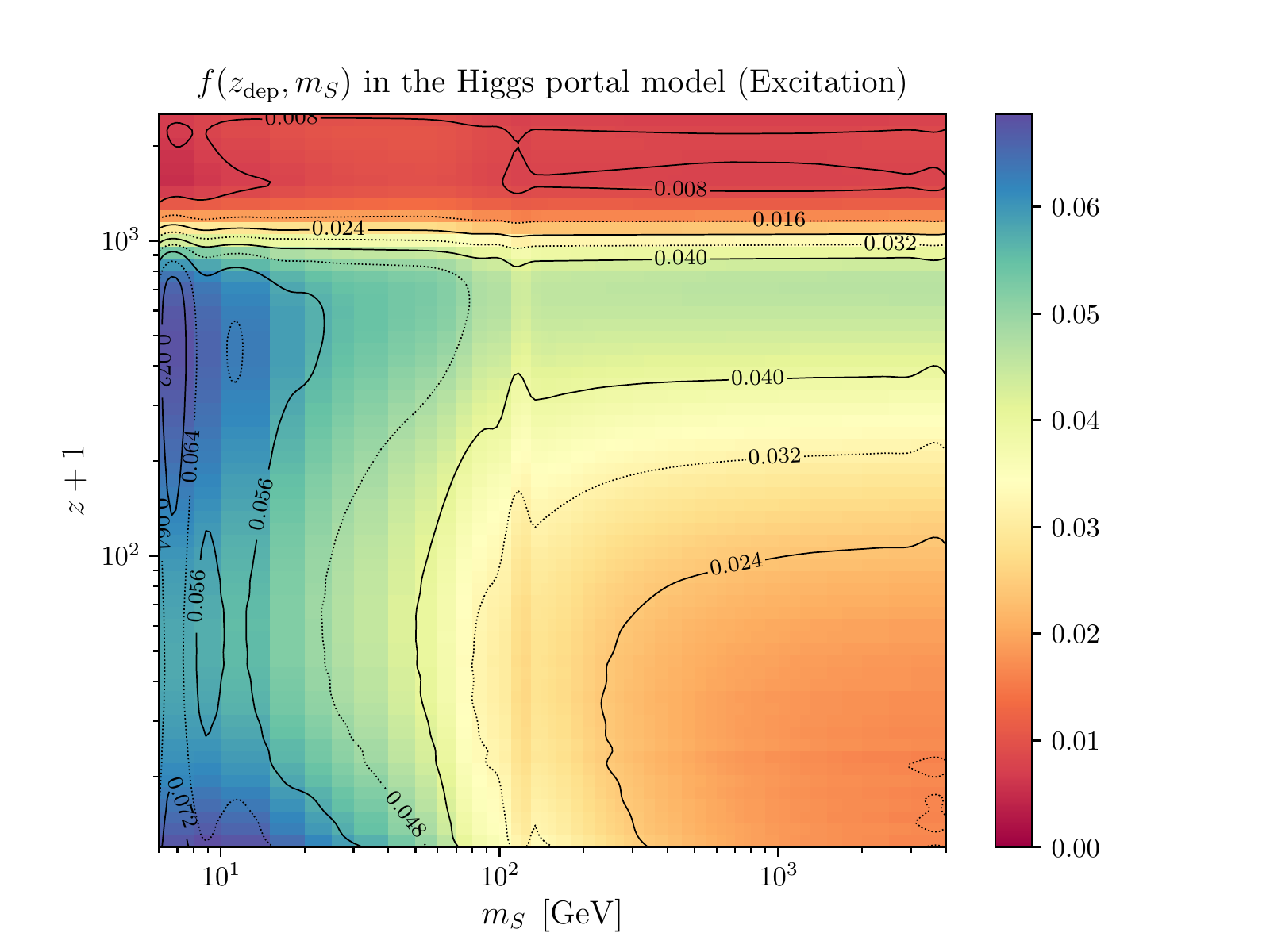}\\
\includegraphics[width=.45\textwidth]{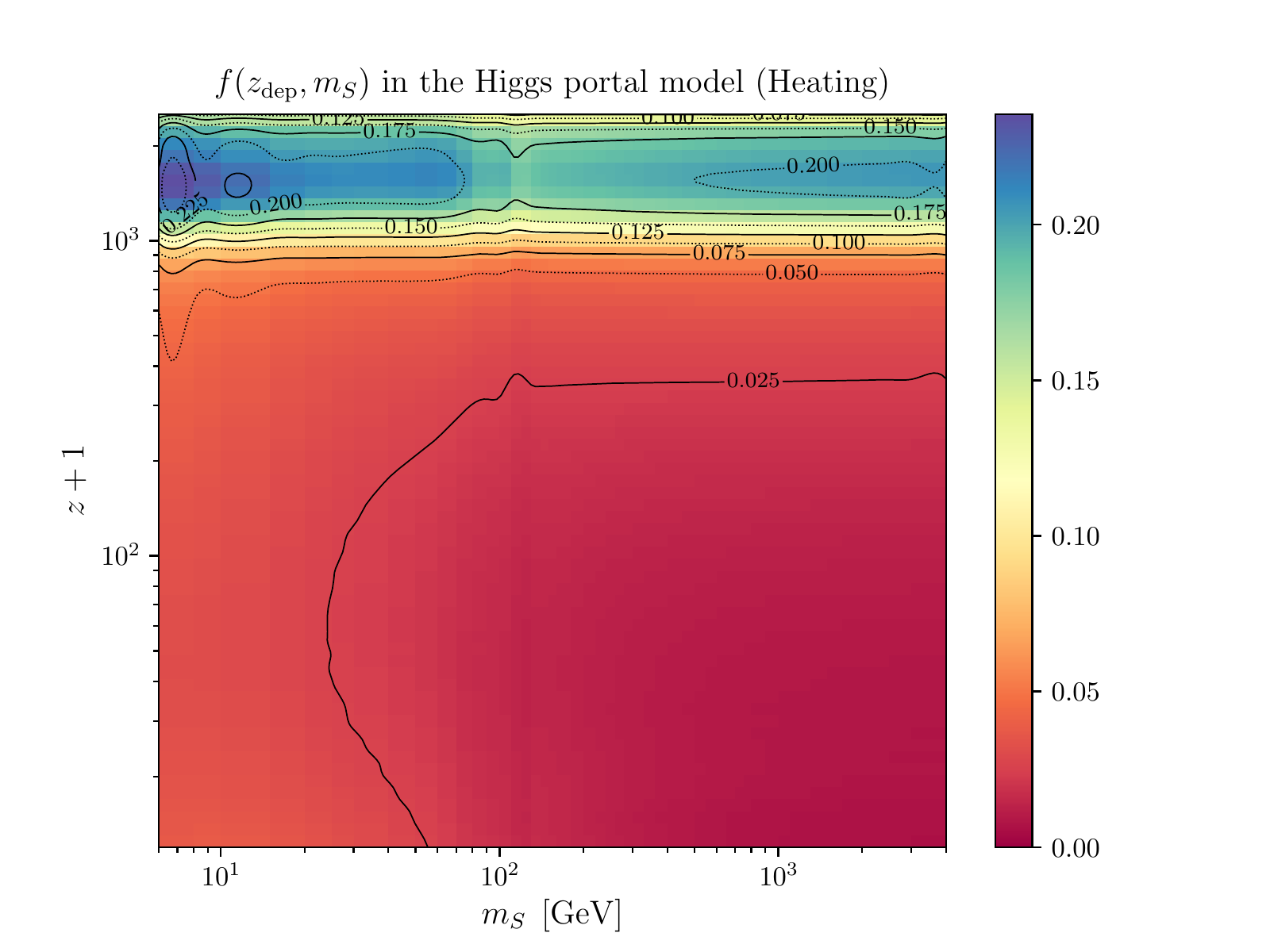}
\includegraphics[width=.45\textwidth]{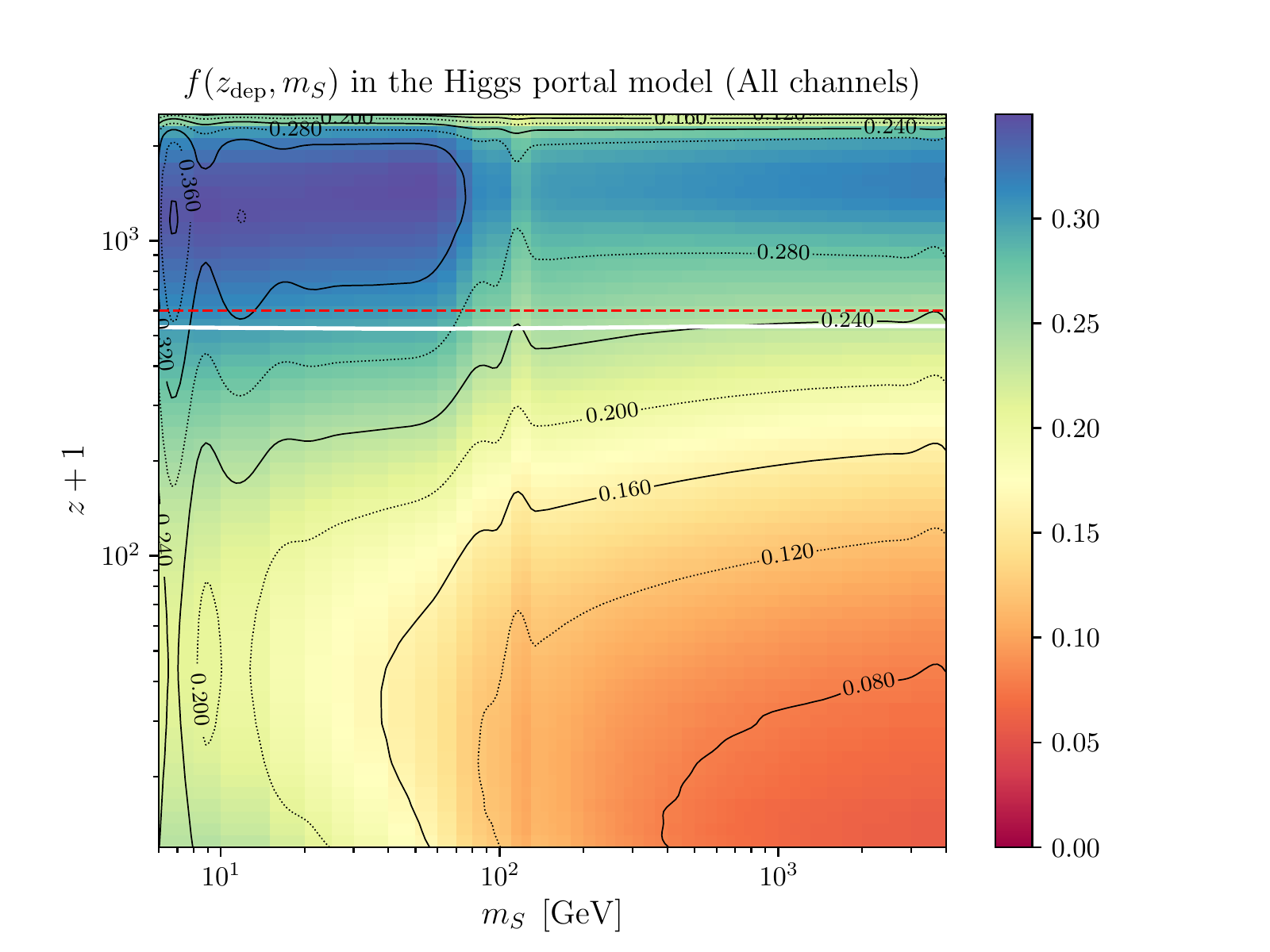}
\caption{\label{fig:HP_fz}Energy deposition functions per channel $f_c(z)$ as a function of the mass in the Higgs portal model .We show the fraction of the injected energy deposited into ionization (top left), excitation (top right), heating (bottom left) and the sum over each channel (bottom right). The dashed red line in the bottom right panel denotes 
the effective redshift $z_{\rm eff}$, defined by $f(z_{\rm eff}) = f_{\rm eff}$, with $f_{\rm eff}$ as proposed in~\cite{Slatyer15-1}. The solid white line denotes $z=600$, which has been used in most previous analyses to evaluate an effective efficiency factor.}
\end{figure}

The total energy deposition function $f(z)=\sum_cf_c(z)$ (equation~\ref{eq:fz_anni}) of the scalar Higgs portal model is presented in figure~\ref{fig:HP_fz}. In the left panel we show a contour plot in the plane of the dark matter mass, $m_S$, and the redshift, $z+1$. The dependence of $f(z)$ on the dark matter mass is small for the large redshifts, $z \gtrsim 500$, which are probed by the CMB analysis. In figure~\ref{fig:HP_fz}, left panel, we also display the effective redshift, $z_{\rm eff}$ (dashed red line), defined by $f(z_{\rm eff}) = f_{\rm eff}$, where $f_{\rm eff}$ is the effective energy deposition factor as proposed in~\cite{Slatyer15-1}. The effective redshift is close to $z=600$ (solid white line) which has been used in most previous analyses to define an effective energy deposition. In the right panel of figure~\ref{fig:HP_fz} we show the $z$ dependence of $f(z)$ for various dark matter masses. 

From our general analysis of the energy deposition function we can conclude that an effective energy deposition factor as proposed in Ref.~\cite{Slatyer15-1} is an excellent approximation to derive CMB constraints on the Higgs portal model. We checked that this conclusion holds also for a future CMB experiments such as CORE+ \cite{DiValentino:2016foa}. Moreover, the effective energy deposition is essentially identical for the different SM final states we consider, i.e.\ annihilation into quarks, gluons, gauge or Higgs bosons, and thus largely independent of the relative size of the various annihilation channels. Thus, the CMB constraints we derive for the Higgs portal model are applicable to a large class of models where direct annihilation into leptons or photons is suppressed. 

\begin{figure}[htb]
\centering
\includegraphics[width=.7\textwidth]{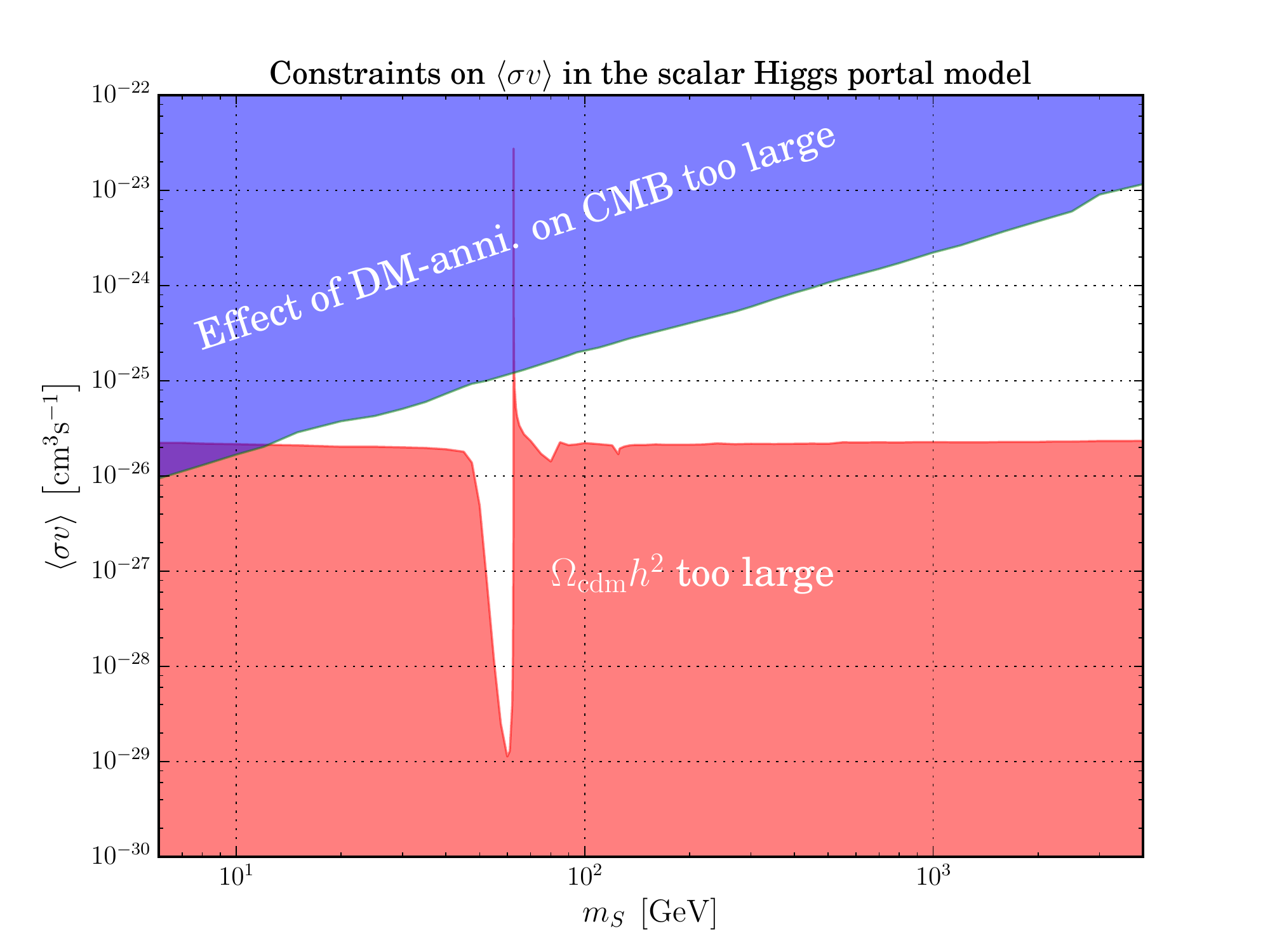}
\caption{\label{fig:HP_constraints}CMB constraints on the annihilation cross-section in the Higgs-Portal model.}
\end{figure}

To constrain the annihilation cross section $\langle\sigma v\rangle$ using Planck data, we use \texttt{MontePython} \cite{Audren2012} and perform a scan over the six $\Lambda$CDM parameters 
\[ \lbrace\Omega_b h^2, \Omega_\mathrm{cdm} h^2, \ln A_s, n_s, z_\mathrm{reio}, H_0 \rbrace\]
and two additional dark matter parameters
\[\lbrace m_\mathrm{DM}, \langle\sigma v\rangle \rbrace. \]
We take flat priors on these parameters, and restrict the
dark matter mass to the range 5~GeV~$< m_S<$~5~TeV. Our data set consists of Planck high-$\ell$ TT,TE,EE + simLOW (prior on $\tau_{\rm reio} = 0.055\pm0.009$) and Planck CMB lensing \cite{Planck2015_13}.  The results are shown in figure~\ref{fig:HP_constraints}. Our analysis excludes large annihilation cross sections, as shown by the green shaded region. We also display the region of parameter space which corresponds to a overabundance of dark matter (red shaded area), which constrains the annihilation cross section from below.

\section{Second application: evaporating Black Holes}
\label{sec:PBH}
\subsection{Energy injection from Hawking radiation}
We now consider a more involved  case, still lacking an accurate treatment, namely the search for evaporating Primordial Black Holes (PBHs). PBHs are created by large density contrasts entering the horizon in the primordial Universe, when the fluid pressure is not able to counteract the gravitational force. PBHs can have a wide range of masses depending on the size of the horizon at collapse, and are often invoked as one of the most appealing alternatives to particle Dark Matter.  Black Holes are expected to experience {\em evaporation}, i.e. loss of mass through the emission of the famous Hawking radiation, predicted by S. Hawking in 1975 for Schwarzschild black holes. The well-known results by Hawking predict a Black Body radiation emitted at a temperature $T_\mathrm{BH}$ depending on the BH mass,
\begin{equation}
T_\mathrm{BH} = \frac{1}{8\pi G M} =  1.06 \cdot \left(\frac{M}{10^{13}\,\mathrm{g}}\right)^{-1} \mathrm{GeV}\,.
\label{eq:bh_mass}
\end{equation}
The  spectrum of particles with spin $ s $ emitted by a black hole with temperature $ T_\mathrm{BH} $ is given by
\begin{equation}
\frac{\diff^2 N_s}{\diff E \, \diff t} = \frac{1}{2\pi^2} \frac{E^2 \sigma_s (M,E)}{\exp{\left(\nicefrac{E}{T_\mathrm{BH}} - (-1)^{2s}\right)}}\,
\label{eq:PBH_spectrum}
\end{equation}  
(see  \cite{Carr2009}),
where $ \sigma_s (M,E) $ is the absorption cross-section. This function has in general a non-trivial form (see \cite{MacGibbon1990}),  but it can be written in the optical (high-energy) limit as
\begin{equation}
\sigma_s (M,E) = 27 \pi G^2 M^2 \quad\text{for $ E \gg T_\mathrm{BH} $}\,.
\label{eq:PBH_crossSection}
\end{equation} 
Given that this function quickly decrease in the opposite regime \cite{PhysRevD.13.198}, we conservatively cut the spectrum at $E=3\,T_{\rm BH}$, just below the peak of the distributions for leptons and photons \cite{Carr2009}. We checked that slighlty varying this threshold doesn't affect our conclusions. 
In the low-mass range, roughly below $10^{17}$g, these BHs are expected to emit particles at energies and rates which can affect the CMB anisotropies, and can thus be looked for in CMB power spectra analysis. This has not been done before with good accuracy, mainly because both the spectrum and the injection rate evolve while the mass decreases. Hence, it is necessary to make use of numerical tools able to track the evolution of both the {\em injected} and {\em deposited} energy rates. This can be done fairly easily with the new \ExoCLASS{} package.
The PBH mass evolves according to \cite{MacGibbon1991}
\begin{equation}
\frac{\diff M}{\diff t} = -5.34\times 10^{25} \mathcal{F}(M) \left(\frac{M}{1\,\mathrm{g}}\right)^{-2} \,\mathrm{\frac{g}{s}}\,,
\label{eq:PBH_massloss}
\end{equation}

where the effective number of degrees of freedom $\mathcal{F}(M)$ is obtained by summing over all possible emitted particles,
\begin{equation}
\mathcal{F}(M) = \sum_{\text{part. $i$}} \Pi_i \cdot f_{s,q} \cdot \exp{\left(- \frac{M}{\beta_s \tilde{M}_i} \right)}\,.
\label{eq:F_of_M}
\end{equation}
In this equation $ \Pi_i $ is the numbers of  internal degrees of freedom of the particles $i$, while the factor $ f_{s,q} $ is an additional weighting factor taking into account the fact that 
particles of different spin $s$ and charge $q$ have slightly different blackbody spectra.

The last two factors in \eqref{eq:F_of_M} are $ \tilde{M}_i $, the mass of a BH when its temperature is equal to the mass of the produced particles $ T_\mathrm{BH} =  m_i $, and $ \beta_s $, which takes into account the shift between the peak of the blackbody distribution and the temperature. 

In table \ref{tab:PBH_F_factors} we give the value of these parameters for each of the particles that we consider. 
\begin{table}[htp]
\centering
	\begin{tabular}{c|ccc|cc}  
	\toprule
	Particles    & $ \Pi_i $ & $ f_{s,q} $ & $ \beta_s $ & $ m_i\,[\mathrm{GeV}] $ & $ \beta_i\tilde{M}_i\,[\mathrm{g}] $ \\
	\midrule
	$ \gamma $ & $ 2 $ & $ 0.060 $ & $ 6.04 $ & $ 0.00 $ & $ \infty $ \\
	$ \nu $ & $ 6 $  & $ 0.142 $ & $ 4.53 $ & $ 0.00 $ & $ \infty $ \\
	$ e $    & $ 4 $    & $ 0.146 $ & $ 4.53 $ & $ 5.11\times10^{-4} $ & $ 9.3969\times10^{16} $ \\
	$ \mu $  & $ 4 $    & $ 0.146 $ & $ 4.53 $ & $ 0.104 $ & $ 4.5429\times10^{14} $ \\
	$ \tau $ & $ 4 $    & $ 0.146 $ & $ 4.53 $ & $ 1.77 $ & $ 2.7022\times10^{13} $ \\
	$ u $   & $ 12 $    & $ 0.146 $ & $ 4.53 $ & $ 2.2\times10^{-3} $ & $ 2.1826\times10^{16} $ \\
	$ d $  & $ 12 $    & $ 0.146 $ & $ 4.53 $ & $ 4.7\times10^{-3} $ & $ 1.0217\times10^{16} $ \\
	$ c $  & $ 12 $    & $ 0.146 $ & $ 4.53 $ & $ 1.28 $ & $ 3.7514\times10^{13} $ \\
	$ s $  & $ 12 $    & $ 0.146 $ & $ 4.53 $ & $ 9.6\times10^{-3} $ & $ 5.0019\times10^{14} $ \\
	$ t $  & $ 12 $    & $ 0.146 $ & $ 4.53 $ & $ 173.1 $ & $ 2.7740\times10^{11} $ \\
	$ b $  & $ 12 $    & $ 0.146 $ & $ 4.53 $ & $ 4.18 $ & $ 1.1488\times10^{13} $ \\
	$ W $  & $ 6 $    & $ 0.060 $ & $ 6.04 $ & $ 80.39 $ & $ 7.9642\times10^{11} $ \\
	$ Z $  & $ 3 $    & $ 0.060 $ & $ 6.04 $ & $ 91.19 $ & $ 7.0209\times10^{11} $ \\
	$ g $  & $ 16 $    & $ 0.060 $ & $ 6.04 $ & $ 0.6 $ & $ 1.0671\times10^{14} $ \\
	$ h $  & $ 1 $    & $ 0.267 $ & $ 2.66 $ & $ 125.1 $ & $ 2.2541\times10^{11} $ \\
	\hline
	$ \pi_0 $  & $ 1 $    & $ 0.267 $ & $ 2.66 $ & $ 0.1350 $ & $ 2.0886\times10^{14} $ \\
	$ \pi^+ $  & $ 2 $    & $ 0.267 $ & $ 2.66 $ & $ 0.1396 $ & $ 2.0198\times10^{14} $ \\
	\bottomrule
	\end{tabular}
\caption{\label{tab:PBH_F_factors} Factors entering the calculation of the effective number of relativistic degrees of freedom $ \mathcal{F}(M) $ of an evaporating black hole, as defined in equation \eqref{eq:F_of_M}. For gluons we take an effective mass of $ m_g \approx 600\,\mathrm{MeV} $ to take confinement into account.}	
\end{table}

\begin{figure}[htb]
\centering
\includegraphics[width=.485\textwidth]{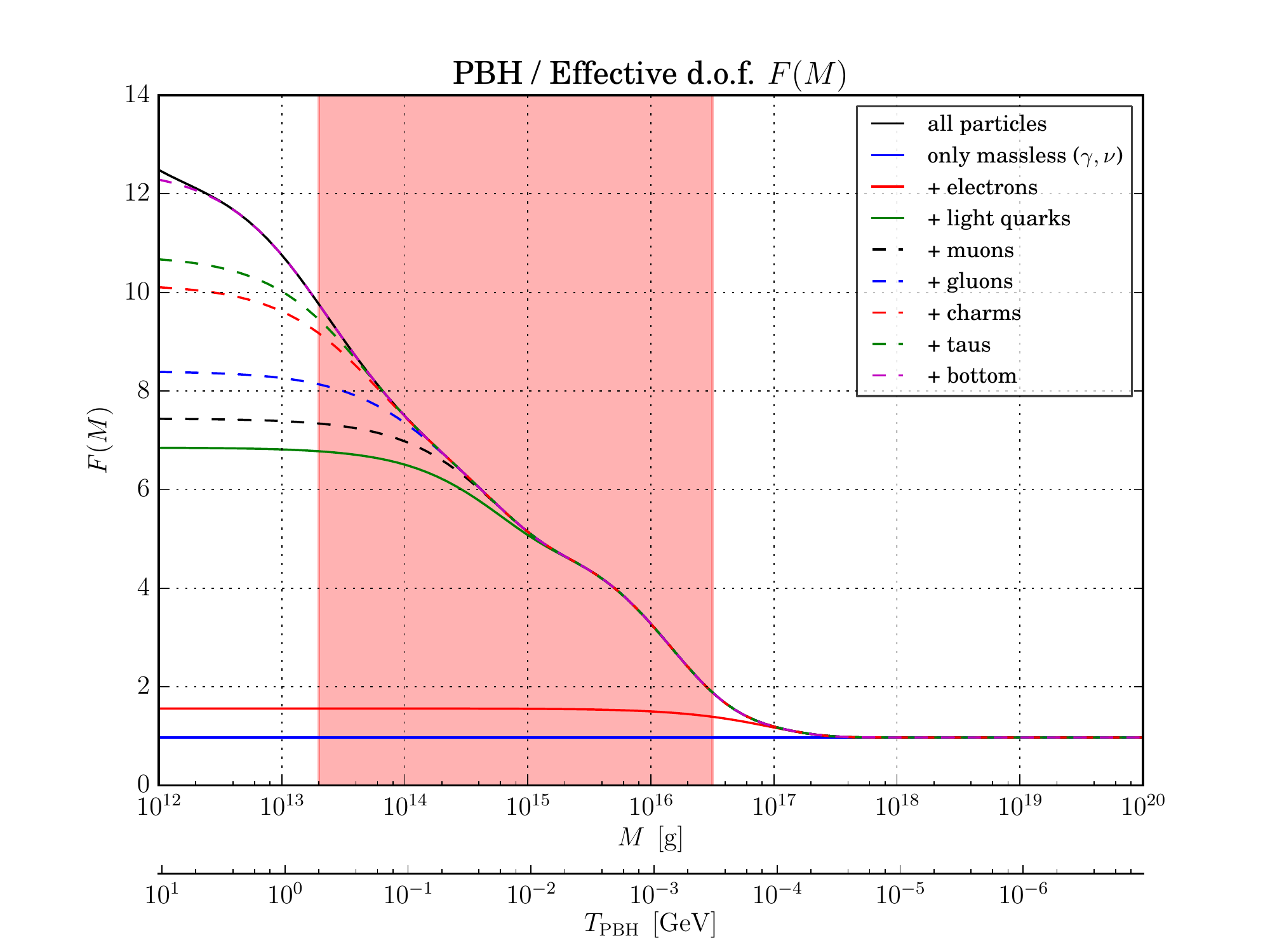}
\hspace{1ex}
\includegraphics[width=.485\textwidth]{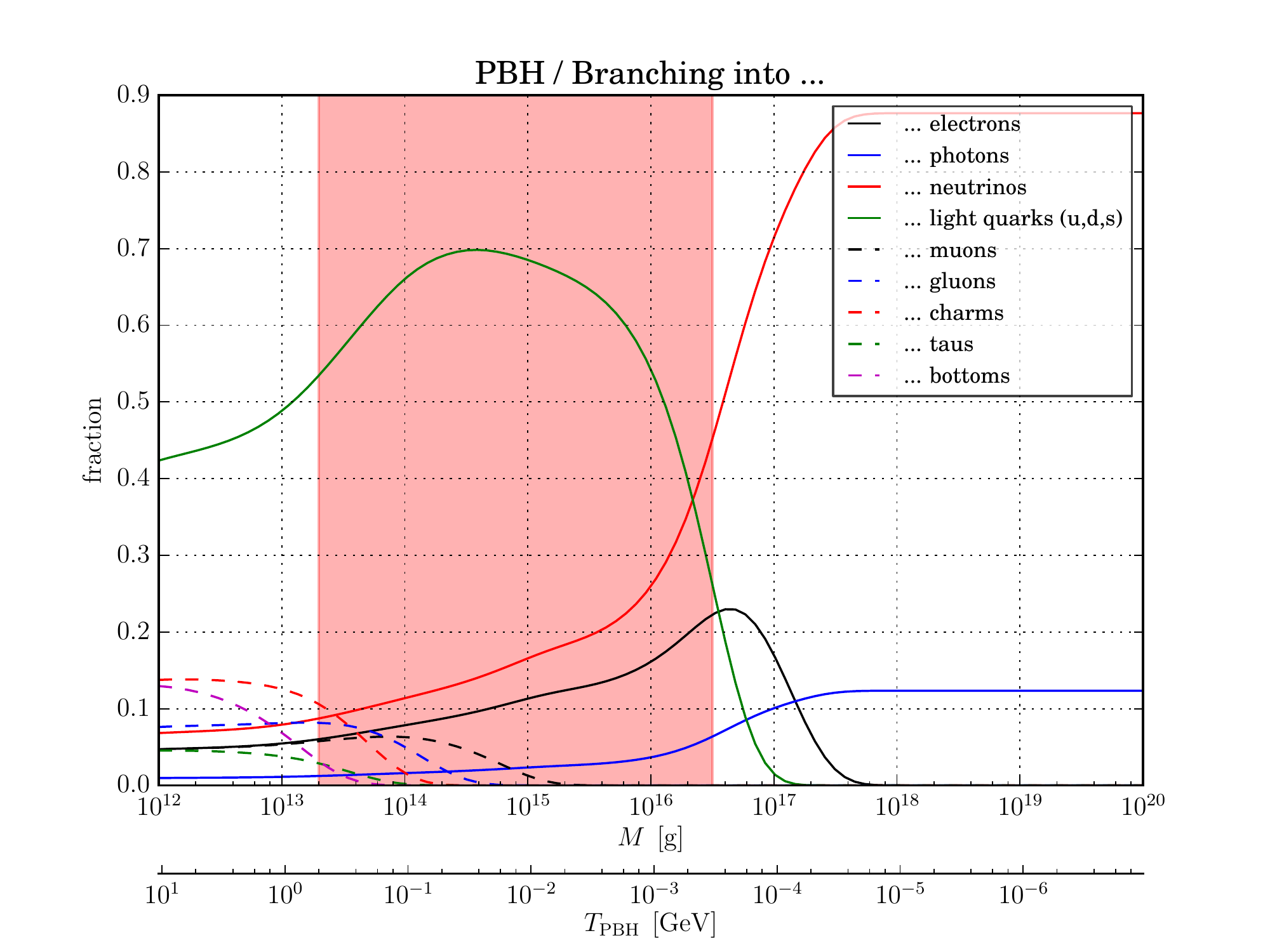}
\caption{\label{fig:PBH_F_and_fraction_noQCD} {\it Left panel:} effective number of relativistic degrees of freedom $\mathcal{F}(M)$. {\it Right panel:} relative contribution of each particle species to according to $\mathcal{F}(M)$.}
\end{figure}

We show $ \mathcal{F}(M) $ and the relative contribution of each particle species to the whole emission in figure \eqref{fig:PBH_F_and_fraction_noQCD}.
We can see that within the mass range of interest, light quarks provide the leading contribution, even in the case where the black hole temperature is around the $ \mathrm{MeV} $ scale. This was overlooked in a previous analysis \cite{Poulin2016}, leading to an overestimate of the bounds in the range $10^{15}$ to $5\times 10^{16}$g.  

One might wonder whether for temperature below the asymptotic freedom limit, BHs radiate fundamental particles or  hadrons. We follow the argument in Refs.~\cite{MacGibbon1990} and \cite{MacGibbon1991}, and assume that a black hole only emit particles that are fundamental at the temperature-scale of the black hole. By this we mean that for temperatures below the QCD-confinement scale (roughly $ \sim 300\,\mathrm{MeV} $), there are no quarks or gluons emitted, but rather pions that decay into photons and electrons. Above this threshold, we assume quarks and gluons to be emitted freely.
In order to take this complication into account, we add to equation \eqref{eq:F_of_M} an exponential suppression factor $ Q_i(T_\mathrm{BH})$ parametrized as
\begin{equation}
Q_i(T_\mathrm{BH}) =
	\begin{cases}
	\left[1 + \exp{\left(-\frac{\log_{10}\left(T_\mathrm{BH} \right)- \log_{10}\left(\lambda_\mathrm{QCD}\right) }{\sigma}\right)}\right]^{-1} & \text{quarks and gluons} \\
	\left[1 + \exp{\left(\frac{\log_{10}\left(T_\mathrm{BH}\right) - \log_{10}\left(\lambda_\mathrm{QCD}\right) }{\sigma}\right)}\right]^{-1} & \text{hadrons}
	\\
	1 & \text{else}
	\end{cases}
\label{eq:QCD_activation}
\end{equation}
where we choose $ \lambda_\mathrm{QCD} = 300\,\mathrm{MeV} $ and $ \sigma=0.1 $.

\begin{figure}[htb]
\centering
\includegraphics[width=.485\textwidth]{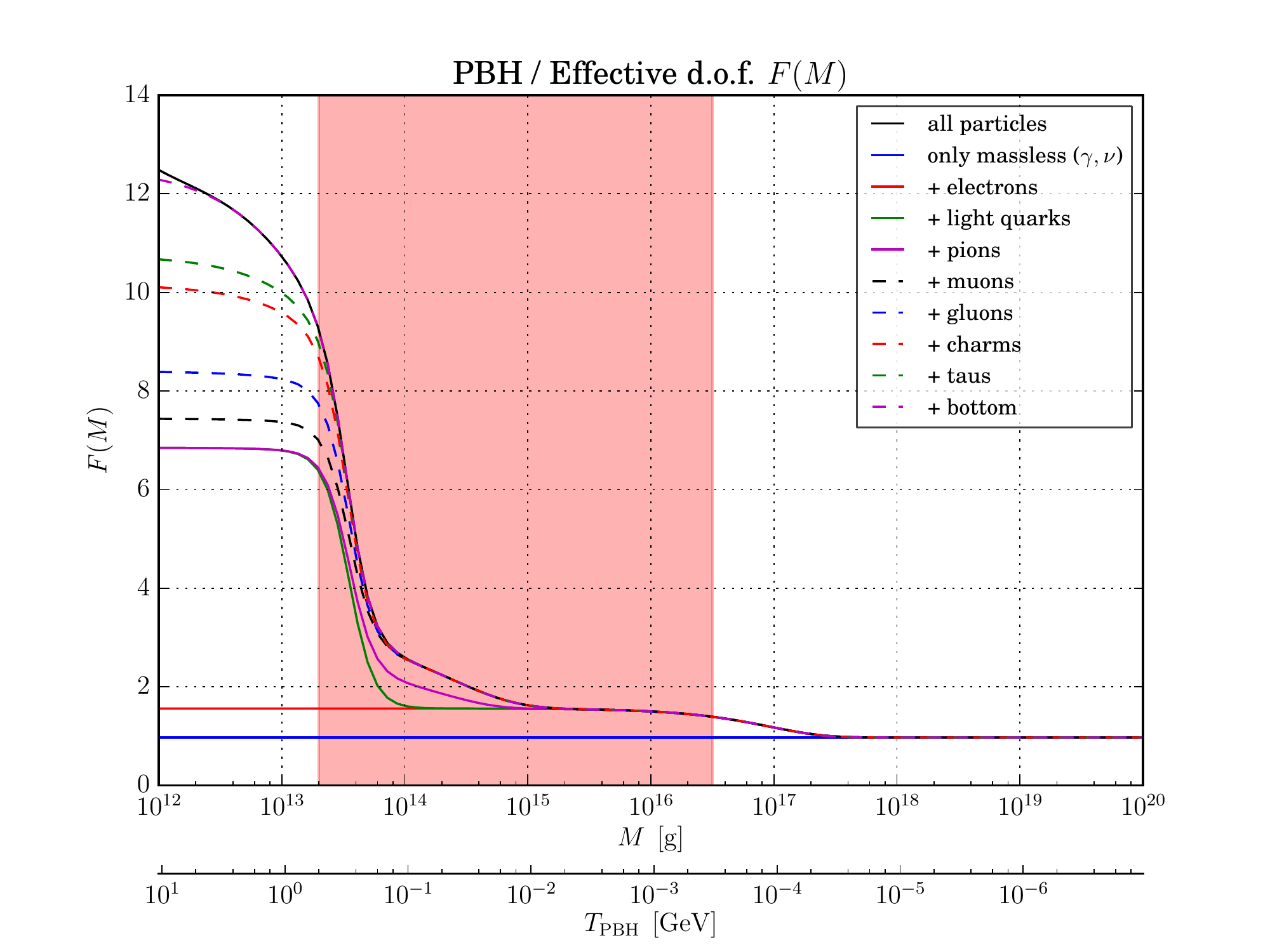}
\hspace{1ex}
\includegraphics[width=.485\textwidth]{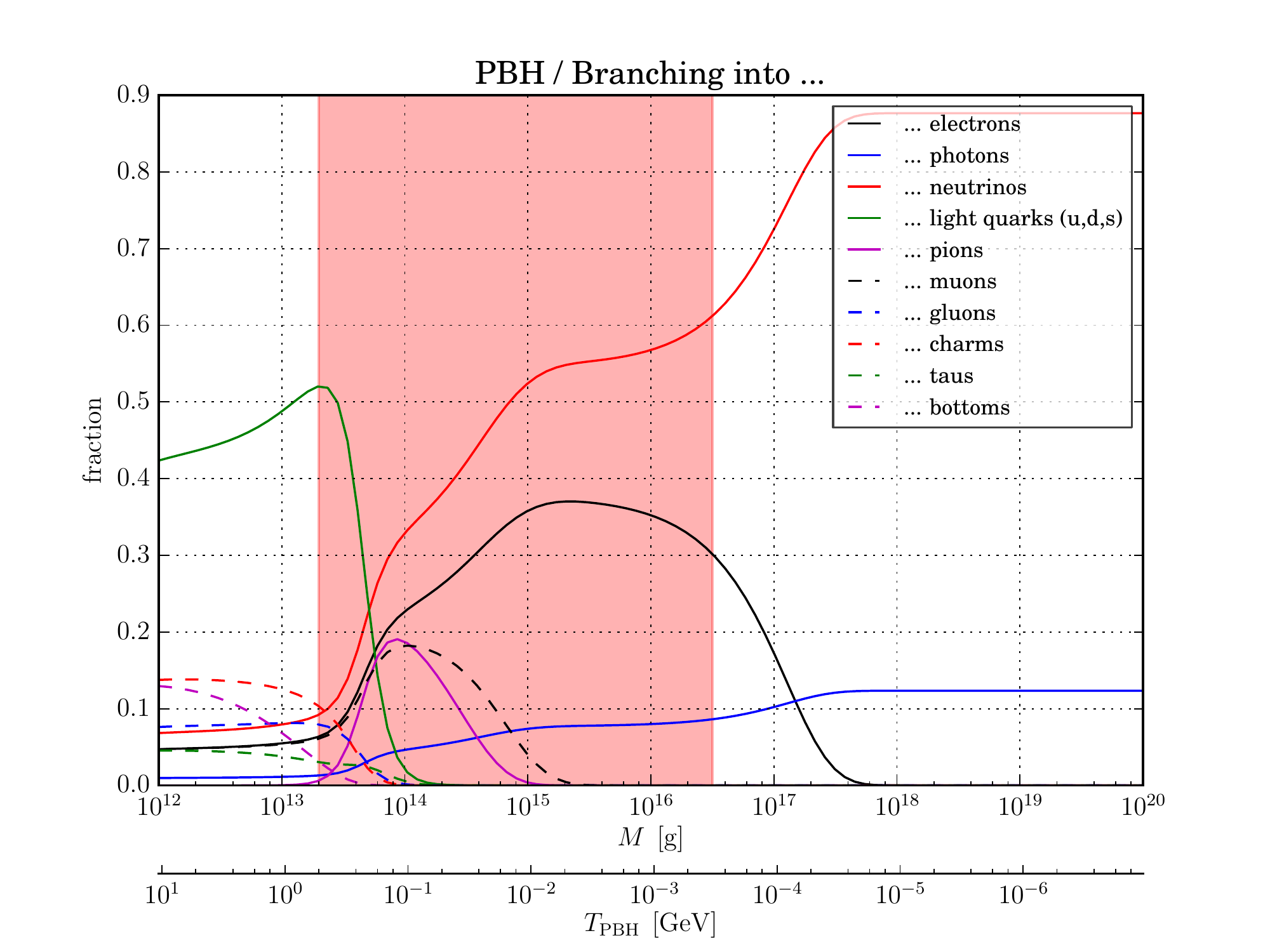}
\caption{\label{fig:PBH_F_and_fraction_QCD}$ \mathcal{F}(M) $ and the relative fraction of each particle species with the inclusion of the QCD-phase transition by the parametrization given in Eq.~\eqref{eq:QCD_activation}.}.
\end{figure}

With this improved parametrization, primary electrons and photons are not strongly suppressed anymore, and the impact of BH evaporation on the CMB is stronger. The emission can mostly be described by electrons, photons, muons and pions (which subsequently decay into photons).  Only for very low masses do quarks become the dominant contribution. The low mass quarks then hadronize and decay, eventually giving rise to photons and electrons. In this study for which particles are injected already at very low energies, it is not possible to make use of the \texttt{PPPC4DMID} spectra, valid only above 5 GeV. For the secondaries of the muons and pions, we thus make use of \texttt{PYTHIA v8.219} \cite{Sjostrand:2014zea}, in which we turn off hadronization/radiative effects. We checked that neglecting these effects do not lead to any sizeable difference on the energy deposition functions even at high energy. This could be expected since these functions measure the calorimetric properties of the plasma. Hence, unlike for cosmic rays searches \cite{Ciafaloni:2010ti}, we don't need to compute the injection spectra in great details.

\begin{figure}[htb]
\centering
\includegraphics[width=.45\textwidth]{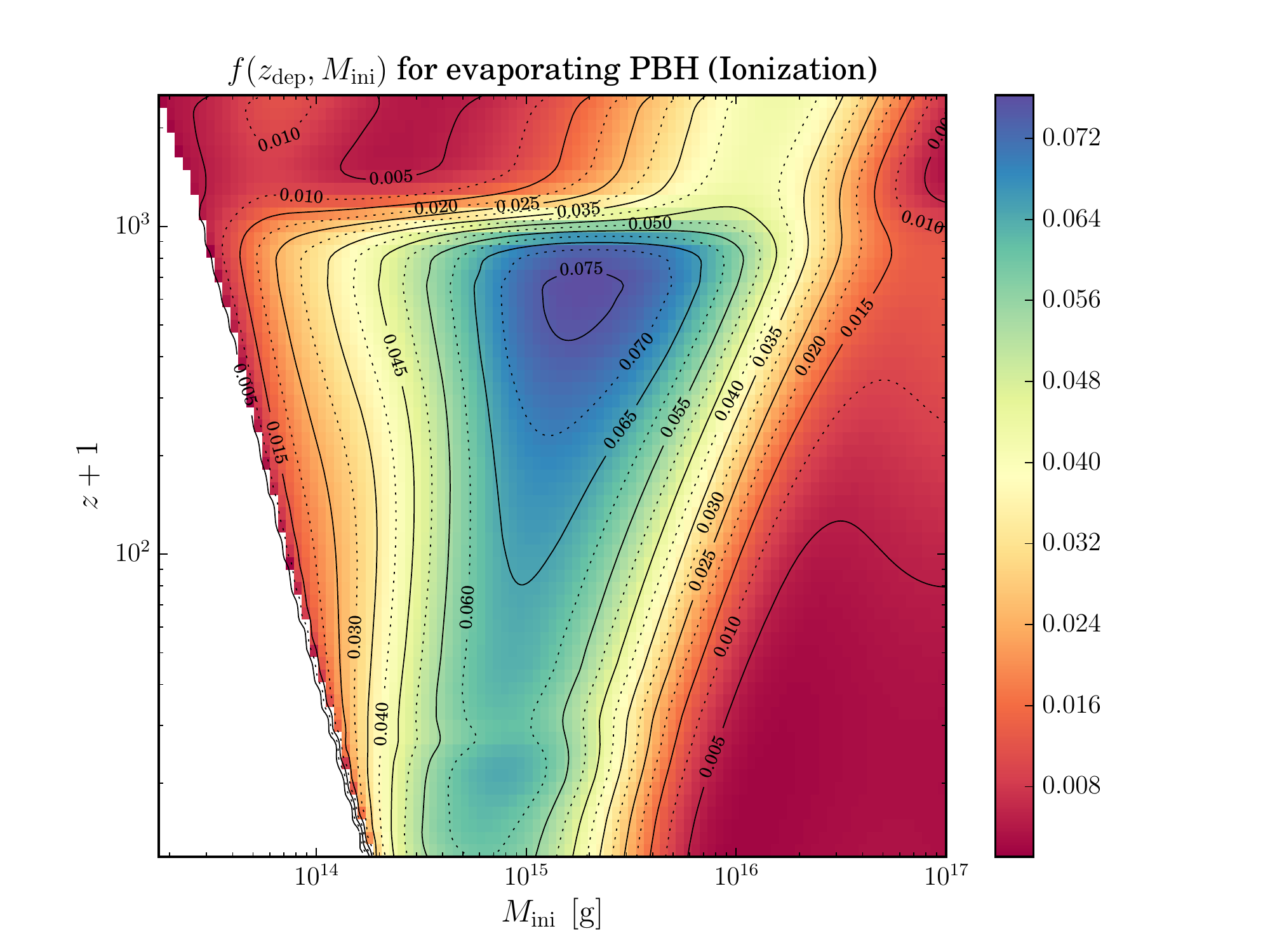}
\includegraphics[width=.45\textwidth]{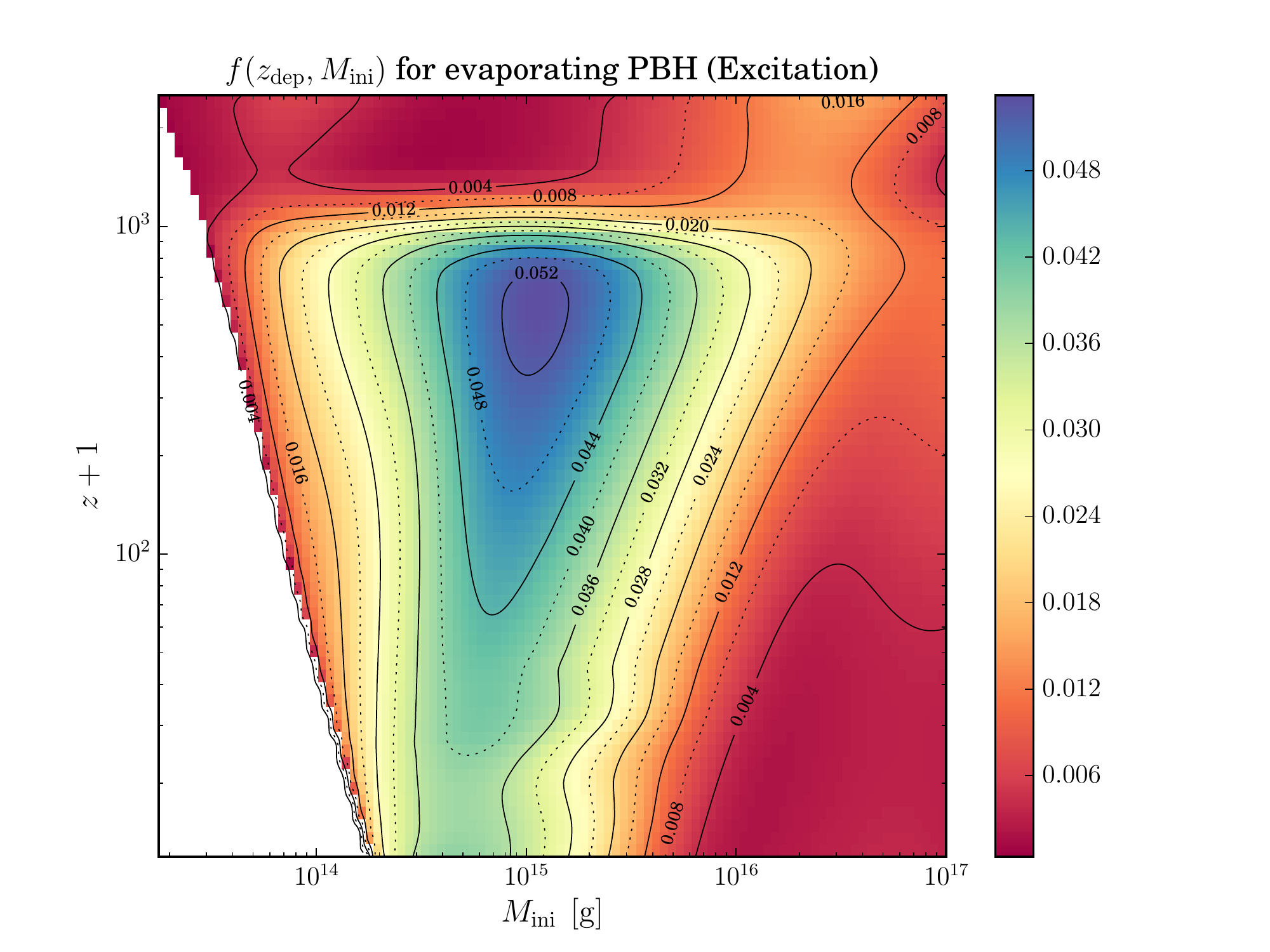}\\
\includegraphics[width=.45\textwidth]{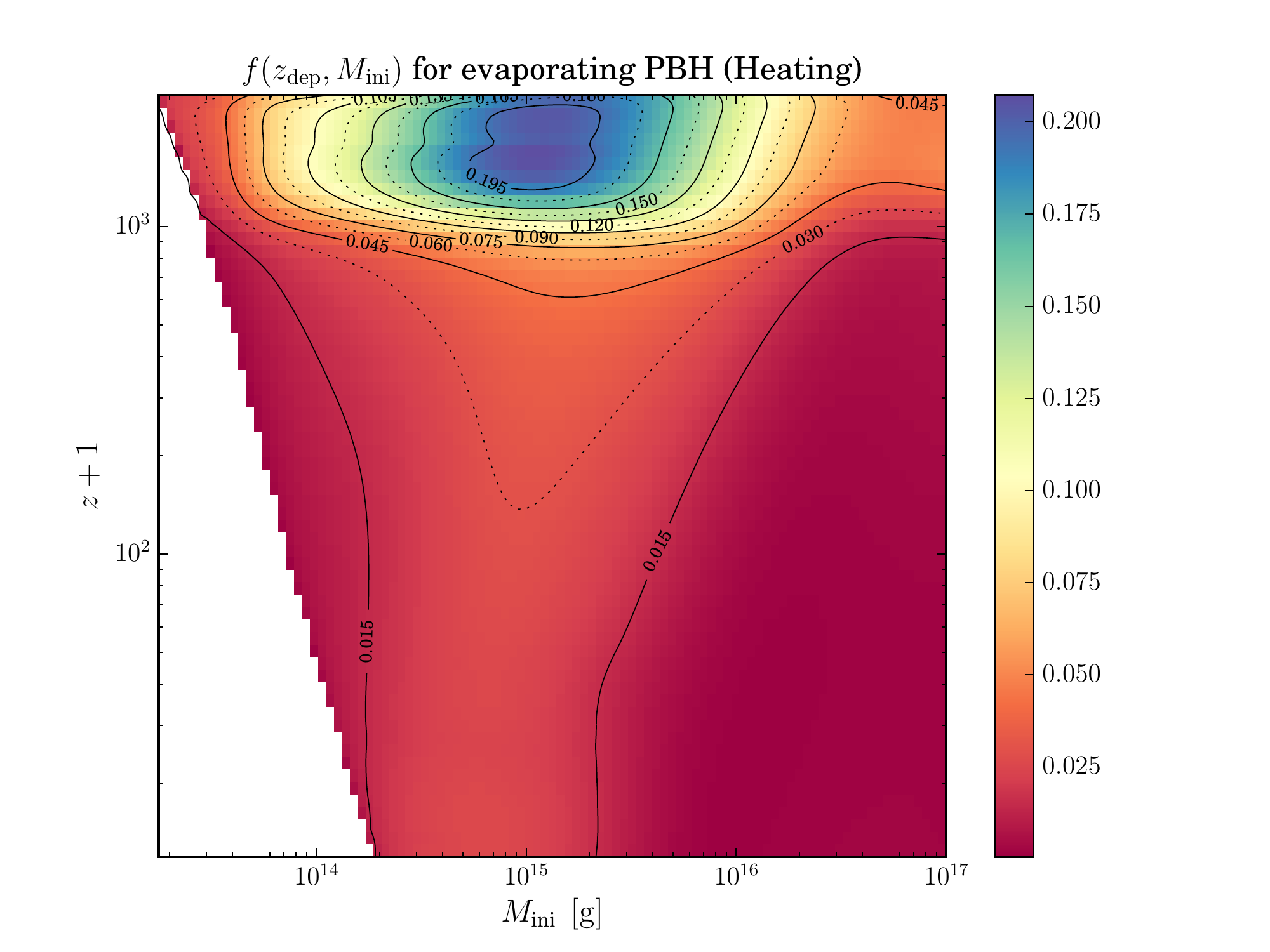}
\includegraphics[width=.45\textwidth]{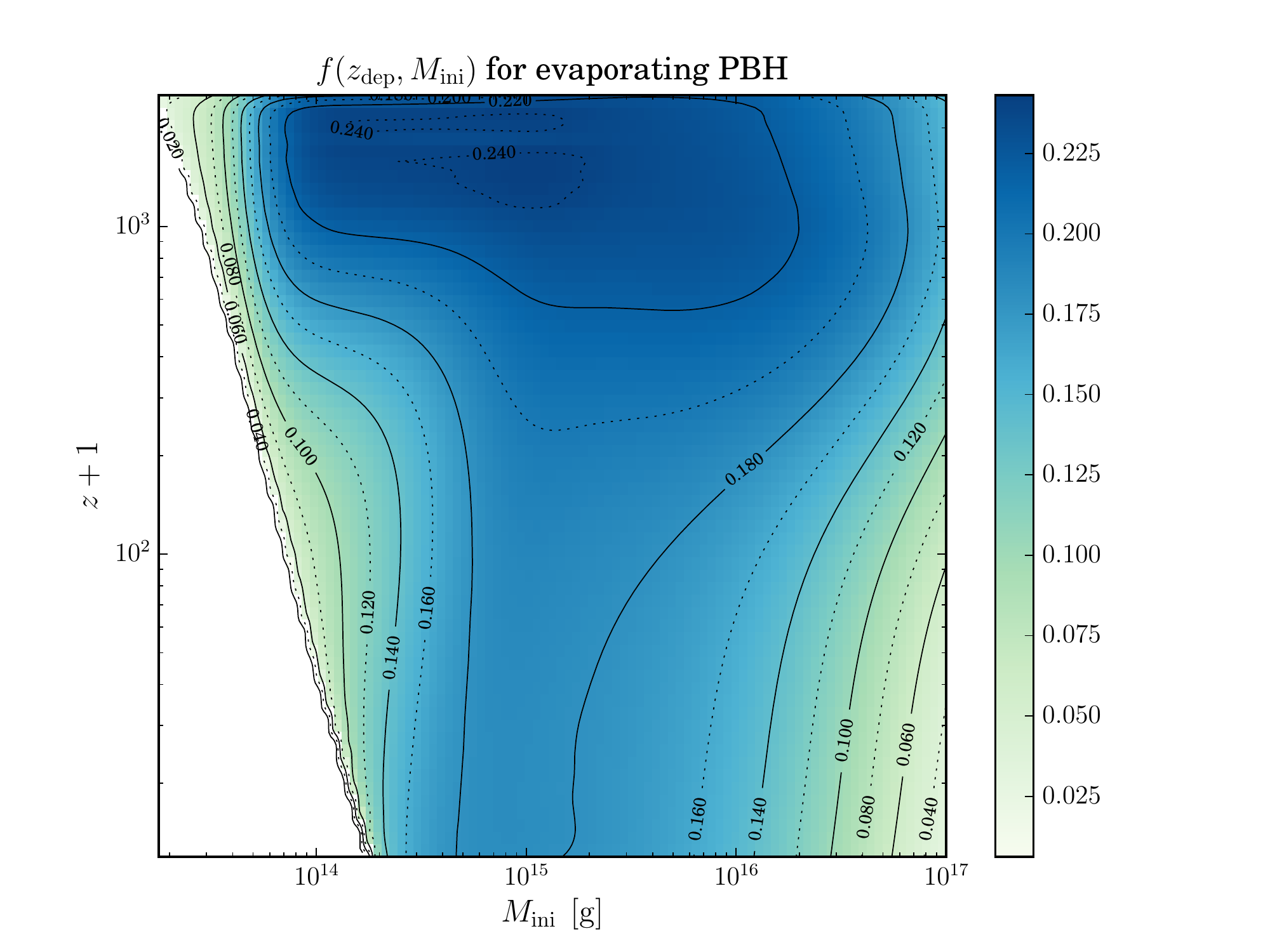}
\caption{\label{fig:PBH_fz_2d} Energy deposition functions per channel $f_c(z)$ for evaporating PBH as a function of the initial mass $ M_\mathrm{ini}$. We show the fraction of the injected energy deposited into ionization (top left), excitation (top right), heating (bottom left) and the sum over each channel (bottom right).}
\end{figure}

We are now ready to compute the relevant energy deposition functions per channel, which we plot in figure~\ref{fig:PBH_fz_2d}.  As discussed in sec.~\ref{sec:2}, the injection rate of electromagnetic energy for low mass PBHs is given by
\begin{equation}
\frac{dE}{dVdt}\bigg|_{\rm inj,~PBH}=\frac{f_{\rm PBH} \, f_{\rm e.m.} \, \rho_c c^2 \Omega_{\rm DM} (1+z)^3}{M_{\rm ini}} \frac{dM}{dt}~.
\label{eq:inj_PBH}
\end{equation}
Here $f_{\rm e.m.}$ is the fraction of the total emitted radiation energy that goes into electromagnetic particles. It is given by  $f_{\rm e.m.} = \mathcal{F}_{\rm e.m.}(M) / \mathcal{F}(M)$, where the sum giving $\mathcal{F}_{\rm e.m.}(M)$ runs over all particles excepted neutrinos. Equivalently, one could omit the $f_{\rm e.m.}$ factor in equation (\ref{eq:inj_PBH}), but evaluate $\frac{dM}{dt}$ from equation (\ref{eq:PBH_massloss}) with neutrinos omitted in the sum of equation (\ref{eq:F_of_M}).

\begin{figure}
\centering
\includegraphics[scale=0.325]{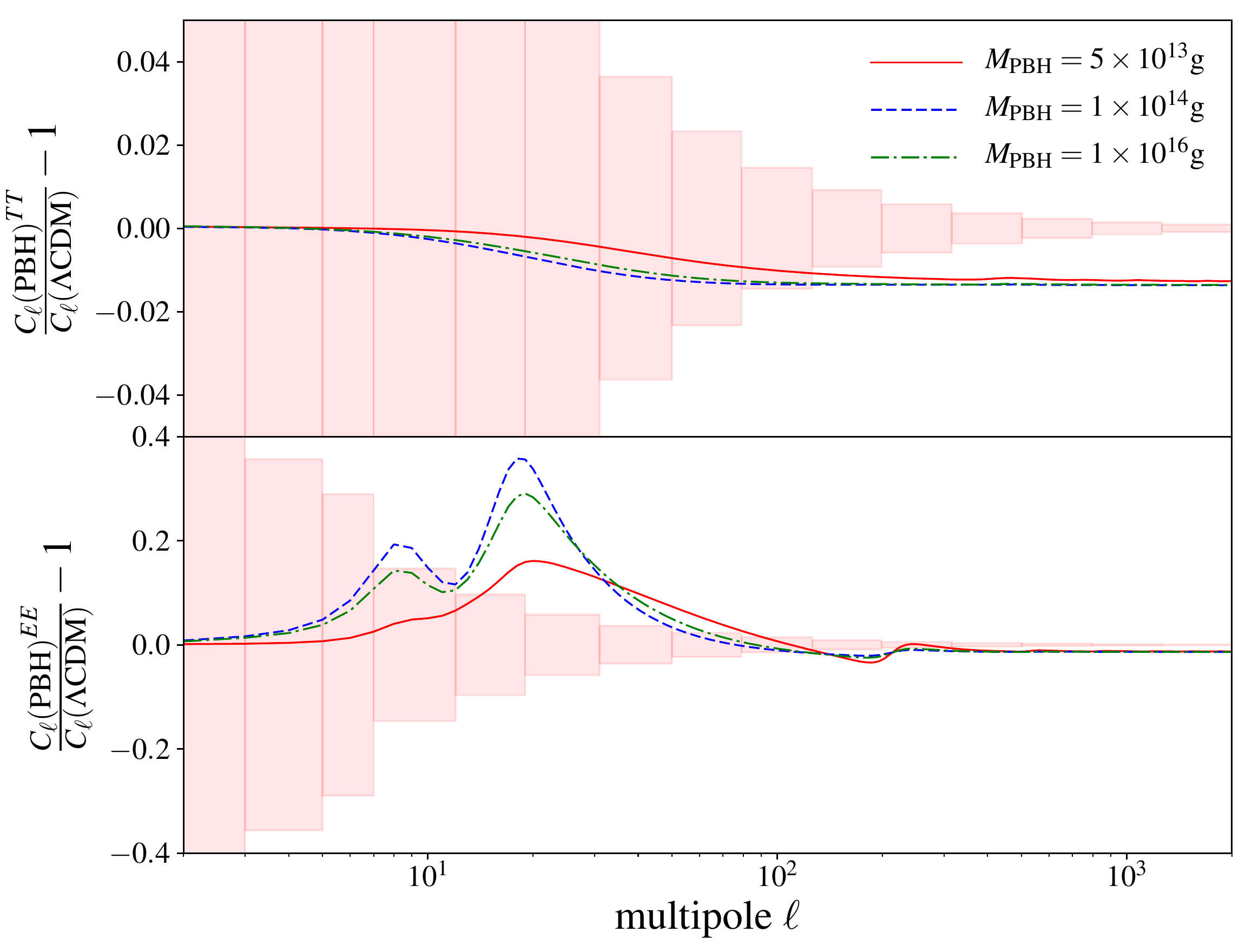}
\includegraphics[scale=0.325]{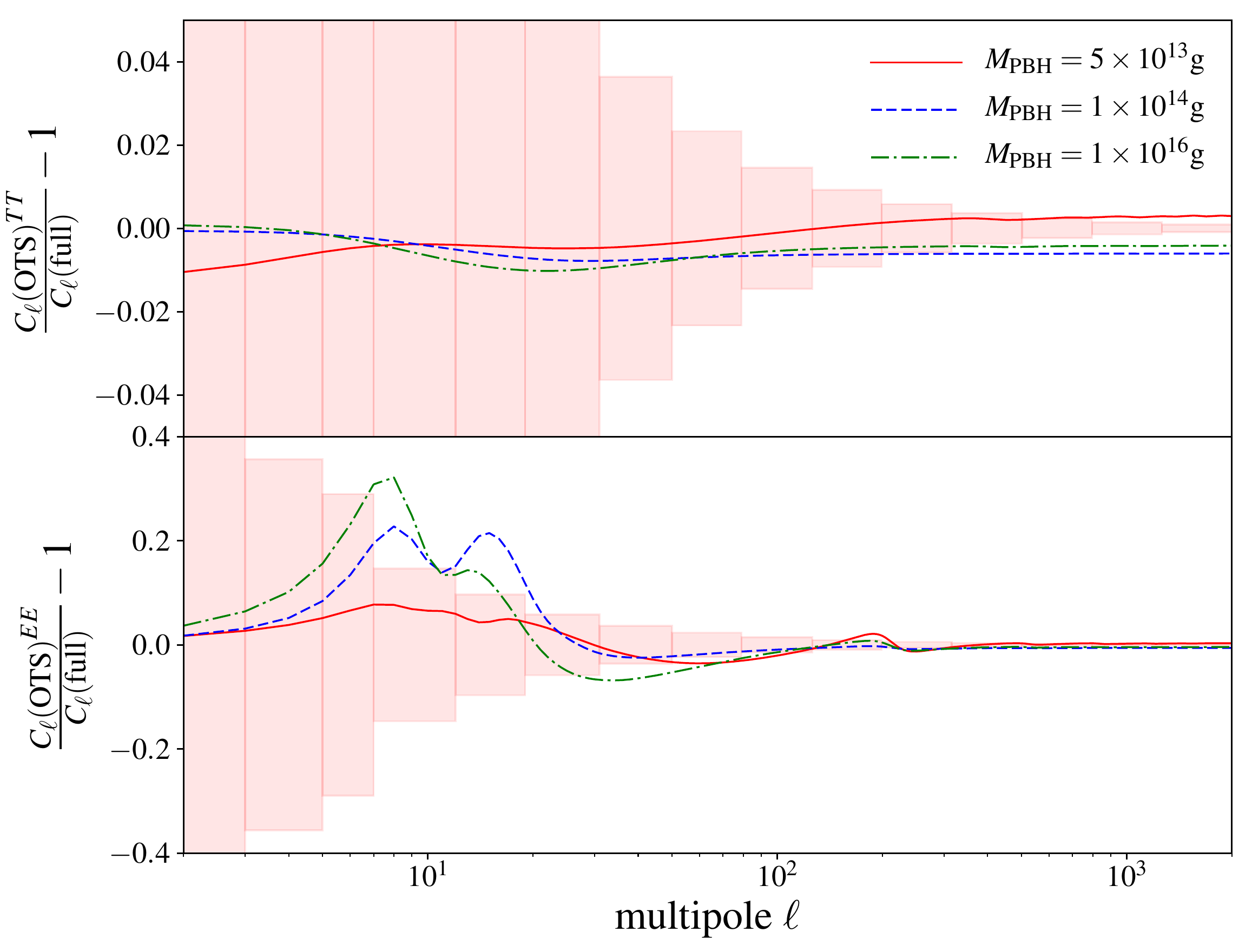}
\caption{{\em Left panel}: Residuals of the unlensed CMB TT and EE power spectra between a universe with evaporating PBH  and the vanilla $\Lambda$CDM with  $\lbrace\Omega_b h^2 = 0.02218, \Omega_\mathrm{cdm} h^2 = 0.1205, \ln (10^{10}A_s)=3.056, n_s = 0.9619, z_\mathrm{reio} = 8.24, H_0 = 67.154 \rbrace$. The density of PBH is set by the constraint at 95\% C.L. obtained in this work and we show $M_{\rm PBH}=5\times10^{13}{\rm g},1\times10^{14}{\rm g}$ and $1\times10^{16}{\rm g}$. {\em Right panel}: Residuals are computed between a model where the impact of the energy deposition is computed with the effective ``on-the-spot'' approximation  and the full computation.}
\label{fig:CMB_PBH}
\end{figure}

We plot on the left panel of figure \ref{fig:CMB_PBH} the residuals of the unlensed CMB TT and EE power spectra, taken between a universe with evaporating PBH  and the vanilla $\Lambda$CDM with  $\lbrace\Omega_b h^2 = 0.02218, \Omega_\mathrm{cdm} h^2 = 0.1205, \ln (10^{10}A_s)=3.056, n_s = 0.9619, z_\mathrm{reio} = 8.24, H_0 = 67.154 \rbrace$. The density of PBH is set by the constraint at 95\% C.L. obtained in this work and we show three different masses $M_{\rm PBH}=5\times10^{13}{\rm g},1\times10^{14}{\rm g}$ and $1\times10^{16}{\rm g}$. The effect of the PBH evaporation and energy injection is typical of any electromagnetic energy injection (see Ref.~\cite{Poulin2016} for a review), although the exact details of the effect depend on the PBH mass. Basically, the increased freeze-out fraction leads to additional Thomson scattering of photons off free electrons along the line-of-sight, which manifests itself as a damping of anisotropies at high-$\ell$'s and an enhanced power in the low-$\ell$ polarization spectrum. For the lowest masses which starts to evaporate earlier, the delayed recombination slightly shifts acoustic peaks and thus generates small wiggles (almost invisble for these tiny PBH fractions) at high multipoles $\ell$ in the residuals with respect to a standard $\Lambda$CDM scenario. 
On the right panel of the same figure, we plot the residuals of the CMB TT and EE power spectra now taken  between a model where the impact of the energy deposition is computed with the effective ``on-the-spot'' approximation as advocated for decaying particles\footnote{Those have a redshift dependence of the injection history close to that of evaporating PBH.} in Refs.~\cite{Finkbeiner11,Slatyer:2016qyl,Poulin2016}, i.e. $f_{\rm eff}=\sum_c f_c(z=300)$, and the full computation. Following the ``corrected 3 keV prescription'' described in Ref.~\cite{Slatyer15-2}, the sum {\em does not} include the lost photon channel. Accordingly, we make use of the energy repartition functions from Ref.~\cite{Galli13} which are called GSVI within \ExoCLASS~(see App.~\ref{sec:ExoClass}). One can see that the result is much bigger than the (binned) cosmic variance represented by the light red boxes.  
We checked that, although it possible to improve the agreement between the approximate spectrum and the full calculation by adapting the value of $f_{\rm eff}$, no unique criterion arises making the use of this approximation much less attractive. More importantly, it is never possible to make the residuals smaller than the cosmic variance. This confirms the need for a more accurate characterization of the signal associated to the energy injection from PBH onto the CMB power spectra (and potentially also for other similar injection histories), and therefore the use of a tool like the \DarkAges~module.

\subsection{Results}
We perform our comparision with CMB data using \ExoCLASS~and \texttt{MontePython} \cite{Audren2012}. We run over the six $\Lambda$CDM parameters plus $f_{\rm PBH}\equiv\Omega_{\rm PBH}/\Omega_{\rm DM}$ (with a flat prior), for 10 different PBH masses distributed between $10 ^{14}\,$ g and $10 ^{16.5}\,$g, as well as 4 masses between $3\times10 ^{13}\,$ g and $10 ^{14}\,$, where a greater accuracy is required. Our data set consists of Planck high-$\ell$ TT,TE,EE + simLOW (prior on $\tau_{\rm reio} = 0.055\pm0.009$) and Planck CMB lensing \cite{Planck2015_13}. 

\begin{figure}
\centering
\includegraphics[scale=0.45]{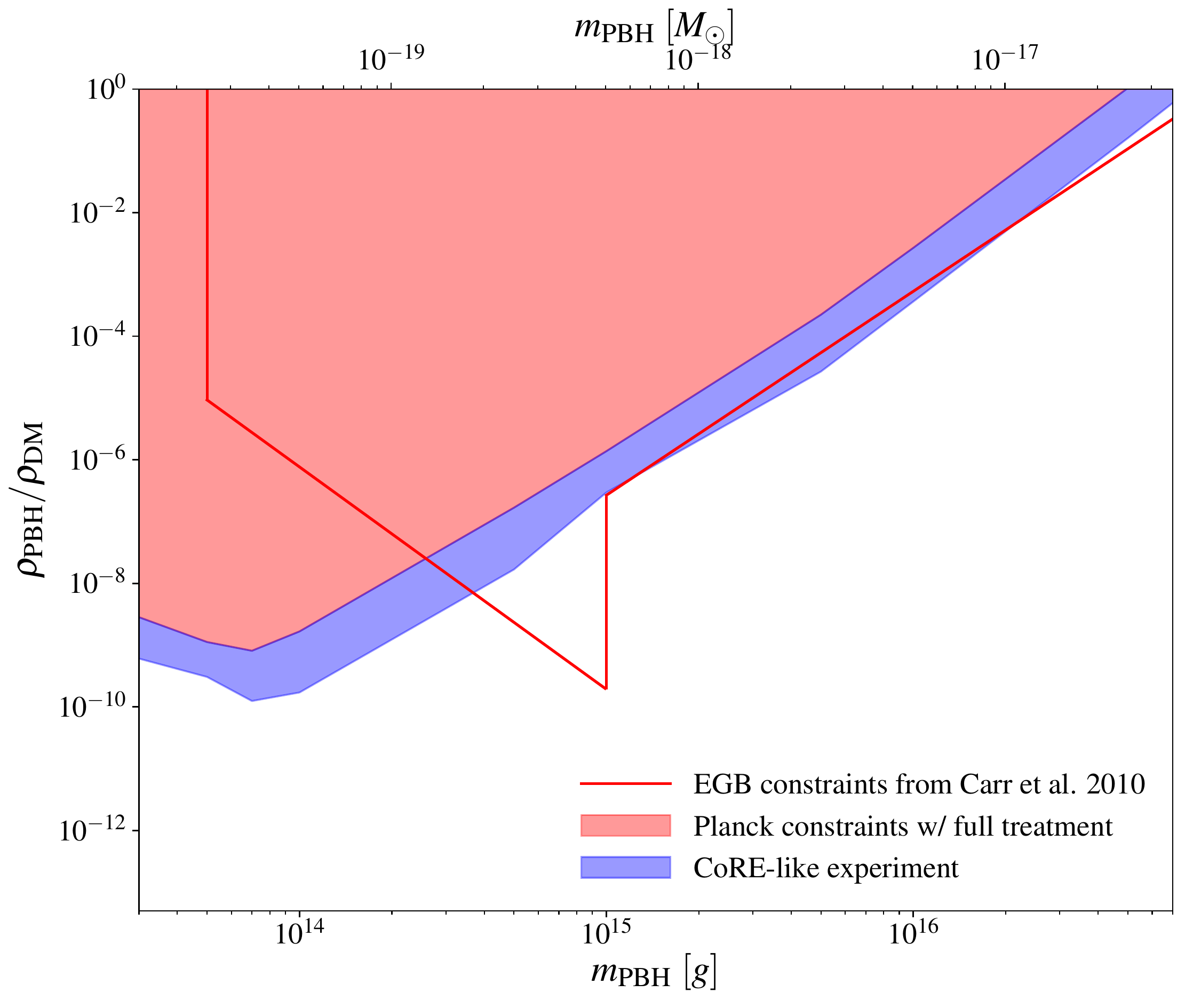}
\caption{CMB and EGB constraints on the fraction of DM made of PBHs in the range $[3\times10 ^{13},7\times10^{16}]$g.}\label{fig:constraints_pbh}
\end{figure}

The results of the MCMC scan in the plane $\{f_{\rm PBH},M_{\rm PBH}\}$ are shown in figure~\ref{fig:constraints_pbh}, together with constraints coming from the Extragalactic Gamma-ray Background (EGB) from Ref.~\cite{Carr:2009jm}. Remarkably, our constraints largely dominate in the range $3\times10^{13}\,$g to $3\times10^{14}\,$g. They are only slightly stronger than the approximate estimate from Ref.~\cite{Poulin2016}. On the other hand, they are a factor of a few weaker than the EGB ones in the range $10^{15}\,$g to $10^{16.5}\,$g.  Note that, although a bit rough\footnote{we estimate $2.5\times10^{13}$g to be a more accurate cutoff, with a very fast loosening of the bound below $3\times10^{13}\,$g.}, the cutoff below $3\times10^{13}\,$g is physical. Below that mass, most of the energy is dissipated in an already fully ionised plasma, and thus the evaporation has no impact on CMB anisotropies. It could however potentially alter BBN and create spectral distortions of the CMB. We do not report these bounds since they go beyond the scope of this paper, but we refer to Ref.~\cite{Poulin2016} for an estimate of the constraining power of these probes compared to CMB anisotropies.

We also perform the same analysis for a future CMB experiment whose specifications are identical to the recently proposed CORE+ experiment with a sky coverage $f_{\rm sky} = 0.70$ \cite{DiValentino:2016foa}. The results are shown as the blue band. One can see that, as found in similar analyses in the context of DM decay \cite{Poulin2016}, up to a one order of magnitude improvement on the sensitivity to $f_{\rm PBH}$ is expected. Interestingly, the CMB constraints would then be comparable to current EGB constraints even in the highest mass range.

\section{Conclusions}

In this paper, we have developed and released a tool allowing one to study the signatures and constraints associated to virtually any form of exotic electromagnetic energy injection. The key quantity to estimate in order to get a reliable prediction of the impact of the energy injection onto the CMB is the energy deposition function per channel. We made use of the results from Ref.~\cite{Slatyer09,Slatyer15-2} obtained in the context of WIMP annihilation, and generalized them to study injection with arbitrary spectra of particles at a given rate. 
Together with this paper, we release the new  \ExoCLASS~branch of the Boltzmann code \CLASS. The new branch already incorporates DM annihilations (including the impact of halos \cite{Poulin2015}), decays from a fraction of DM \cite{Poulin2016}, PBH evaporation and accretion (both spherical and disk \cite{Poulin:2017bwe}). The \ExoCLASS branch involves a new module called \DarkAges, which is easy to modify in order to incorporate any alternative history that one would want to study. Hence, it makes it a very useful tool for the search of DM particle(s) beyond the WIMP paradigm.

We have applied our tools to two particular examples which had not been accurately analysed before: the scalar Higgs portal and low mass ($M\lesssim10^{17}$g) evaporating PBHs. 
The first study allowed us to validate our tool against the standard ``effective on-the-spot'' approximation.  We have derived the most up-to-date cosmological constraints on the scalar mass and Higgs portal coupling to the SM: when requiring the relic density not to overshoot the current measurements, we can identify the surviving part of the parameter space which will partly be probed by the next generation of CMB experiments. Note that the limits derived from energy injection are rather generic and hold for a wide class of models where dark matter annihilates predominantly into gluons, quarks, gauge or Higgs bosons. In the second study concerning evaporating PBHs, we have shown that CMB constraints largely dominate in the range $3\times10^{13}\,$g to $2.5\times10^{14}\,$g, validating the results of Ref.~\cite{Poulin2016}. However, they are slightly weaker than EGB constraints above these masses. Still, this represents an independent cosmological constraints on the presence of evaporating PBHs, which rules out the possibility for them to comprise a large fraction of the DM for PBHs with monochromatic distribution of masses in the range $[3\times10^{13},2.5\times10^{16}]$g. A CMB next-generation experiment such as CORE+ on the other hand could provide constraints competitive with the EGB ones.

The  \ExoCLASS~branch represents a step toward a more accurate characterisation of the impact of a broad range of DM models onto the reionization history and CMB power spectra. This is a key aspect in order to precisely measure properties of the particle or astrophysical DM candidate, if one day a signal was to be found. The next step in this characterization will be to improve the accuracy of the energy deposition tool and interface with the recombination history, today valid at the $\sim20\%$ level \cite{Galli13}. Also, it would be interesting and straightforward to use our tools to study the impact of energy injection on the 21cm line from the epoch of reionization and the Dark Ages, a channel known to be extremely sensitive to these scenarios \cite{Poulin2016}. Moreover, it is important to get our tools ready given that numerous experiments such as PAPER 64~\footnote{\url{http://eor.berkeley.edu}}, 21CMA~\footnote{\url{http://21cma.bao.ac.cn}}, MWA~\footnote{\url{http://www.mwatelescope.org}}, LOFAR~\footnote{\url{http://www.lofar.org}}, HERA~\footnote{\url{http://reionization.org}} or SKA~\footnote{\url{http://www.skatelescope.org}}, are now  (or will be) attempting at measuring this signal.

\label{sec:conclu}
\acknowledgments

We thank warmly Pasquale D. Serpico for very useful discussions and comments. We thank Jan Heisig for providing us the standard model branching ratio of the scalar Higgs Portal model.  This work has been partly done thanks to the facilities offered by the Universit\'e Savoie Mont Blanc MUST computing center.
Parts of the simulations were performed with computing resources granted by RWTH Aachen University under project thes0264. PS is supported by the DFG Emmy Noether Grant No. KA 4662/1-1.

\appendix
\section{Description and usage as a stand-alone package}
\label{sec:DarkAges}

Within the package folder, any call to the \DarkAges~script can be made by simply typing\footnote{ Within the \CLASS~folder the exact command will be \texttt{./DarkAgesModule/bin/DarkAges}.} \verb|./bin/DarkAges|. To specify the spectra and the injection history, the script can be called with additional arguments. These are:
\begin{description}
 \item{\verb|--hist|} This option is used to specify the injection history, as described in previous section. The valid options so far are \verb|annihilation| for $s$-wave annihilation, \verb|annihilation_halos| for $s$-wave annihilation with the impact of the halo boost $\mathcal{B}(z)$, \verb|decay| for a decaying dark matter component with constant lifetime, \verb|evaporating_PBH| for light primordial black holes, and \verb|accreting_PBH| for heavy primordial black holes.\\
 Depending on the history chosen, additional options need to be set.
 \begin{itemize}
  \item For the case of decaying dark matter (\verb|--hist=decay|) the lifetime $\tau$ (in seconds) needs to be set with \verb|--tdec|.
  \item For the case of halo boosted dark matter annihilation (\verb|--hist=annihilation_halos|) the parameters $f_h$ and $z_h$ for the parameterization of $\mathcal{B}(z)$ (equation (C.5) of \cite{Poulin:2015pna}) need to be set with \verb|--zh| and \verb|--fh|)\,.
  \item For heavy primordial black holes (\verb|--hist=accreting_PBH|), the accretion recipe used to compute the energy injection needs to be specified with \verb|--accretion_recipe|. The options to choose are \verb|spherical_accretion| or \verb|disk_accretion|.
 \end{itemize}
 \item{\verb|-m / --mass|} This is the mass of the dark matter candidate. For the scenarios included so far the units are $\mathrm{GeV}$ for annihilating and decaying dark matter, $\mathrm{g}$ for light (evaporating) primordial black holes, and $M_\odot$ for heavy primordial black holes.\\
 Alternatively, it is also possible to specify the logarithm to base 10 of the mass with the option \verb|--log10mass|.

 \item{\verb|-s / --spectrum|} In scenarios in which the injection spectra are not automatically set by the model,
 
 the injected spectrum of electrons, positrons and photons need to be given.  As of now, the scenarios where this is needed are \verb|annihilation|, \verb|annihilation_halos|, and \verb|decay|. The code can work with multiple spectra which are given as a list, separated by blanks. There are three possibilities to pass the spectrum.
 \begin{itemize}
  \item If the spectra of electrons/positrons and photons are stored in a file, the path to this file can be used as an input for the spectrum. In its default setting the code assumes the file to contain the following table,
  \begin{center}
  \begin{tabular}{c|c|c|c|c}
  \toprule
  $ m_\mathrm{DM} $ & $ E $ &  $ \left.\frac{\diff N}{\diff E}\right\vert_{e^\pm} $ & $ \left.\frac{\diff N}{\diff E}\right\vert_{\gamma} $ & $ \left.\frac{\diff N}{\diff E}\right\vert_\mathrm{oth.} $ \\
  \midrule
  \ldots & \ldots & \ldots & \ldots & \ldots \\
  \bottomrule 
  \end{tabular}
  \end{center}  
If more convenient, it is the possible to use $ \log_{10} E $ rather than $ E $ and $ \frac{\diff N}{\diff \log_{10} E} $ rather than $ \frac{\diff N}{\diff E} $, by passing additional keyword arguments to the interpreter. For details we refer to the documentation and the examples provided in the directory of the \DarkAges{}  module.
  If the spectra in this table are given for multiple masses (masses need to be ordered in growing order), the spectra will be automatically interpolated for masses within the given range.
  \item If the input is one of the following keywords (\verb|electron|, \verb|muon|, \verb|tau|, \verb|quark|\footnote{The light quarks $ u,d,s $ are treated as one particle}, \verb|charm|, \verb|bottom|, \verb|top|, \verb|wboson|, \verb|zboson|, \verb|gluon|, \verb|photon|, \verb|higgs|), the code will use the \texttt{PPPC4DMID} spectra \cite{Cirelli2010,Ciafaloni:2010ti} for dark matter masses between $ 5\,\mathrm{GeV} $ and $ 100\,\mathrm{TeV} $.
  \item If the input is either \verb|dirac_electron| or \verb|dirac_photon|, a dirac distribution at the energy given by the dark matter mass (for annihilation) or  half the dark matter mass (for decay) will be used as the spectrum of $ e^\pm $ and $ \gamma $. \footnote{Please note, when using a dirac-like spectrum, it is not yet possible to combine them with spectra of the first two kinds.}.
 \end{itemize}

 \item{\verb|-b / --branching|} When more than one spectrum is used in \verb|--spectrum|, the relative branching ratio between the channels need to be given separated by blanks. The list needs to have the same nuber of entries than \verb|--spectrum| and they need to add up {\em exactly} to one.
\end{description}

Per default, the spectra are sampled at the energies given in the tables of the transfer functions. The user has also the possibilty to give a custom energy range (equally spaced in logarithmic space). The options to be used there are:
\begin{description}
 \item{\verb|--log10Emin|} The lower bound of the energy table (the energy is given in units of $\mathrm{eV}$).
 \item{\verb|--log10Emax|} The upper bound of the custom energy table.
 \item{\verb|--nbins_table|} Number of bins of the custom energy table.
\end{description}
In case the range asked by the user goes beyond the energy range sampled by the transfer functions, it is currently assumed that these are constant, equal to the very lowest and very highest point of the energy grid. As discussed in section~\ref{sec:2}, this is known to be a very good approximation for the lowest energies (up to $\sim100$ eV, we conservatively set the transfer function to zero below). Inspection of the transfer functions at the highest energies show that it is also fairly good for these, although a dedicated study would be needed to safely go above the hundreds of TeV scale.

This concludes the use of the basic usage of the \DarkAges~module. However, when analyzing particular dark matter models, the preferable choice of inputs are often parameters of the model (e.g. masses of the dark matter particles and couplings) rather than the use of \verb|--branching| and \verb|--specfile| as inputs for every parameter point. For that reason we include the possibility to define custom \emph{physics models} which translate the custom model parameters into given spectra and branching ratios.  In that case, the user needs to provide the spectra and branching ratios {\em on a grid of the model parameters} which will later be interpolated. 
In order to reduce the execution time of the code, this grid is read once in a {\em preparation  step} at which the interpolation tables are initiated and saved. Later on, the {\em execution  step} is performed and the spectra and branching ratios are sampled at the required point in parameter space.\\
This mode can be invoked via \verb|--model=MODEL|. Here, \verb|MODEL| defines the name of the folder where the routines of the physics model are located (\verb|./models/MODEL/|). Inside this folder the code expects at least one Python  script which defines the following routines 
\begin{description}
 \item{\verb|prepare()|} This routine reads the branching ratios and spectra given on a grid of the model parameters and initiates the interpolations and saves them. This routine will only be run once, when \DarkAges{} accesses the model for the first time. At a later point this step can only be accessed by forcing it with using the option \verb|--rebuild-model|.
 \item{\verb|run(*arguments, **DarkOptions)|} This routine takes the interpolations of the spectra and branching ratios performed and saved by the \verb|prepare()|-routine and samples them at the model parameters. Here, \verb|*arguments| is a list of the custom model parameters and \verb|**DarkOptions| are additional keyword  arguements, such as precision parameters or handles for non-default behaviour of the functions within \DarkAges{}.
\end{description}
For more details we refer the reader to the documentation and to the ``toy  model'' located at \verb|./models/simple_mix|.

To pass additional options to the functions of the \DarkAges~module, most functions take additional keyword  arguemts which are located within the globally accessible dictionary \verb|DarkOptions|. The option \verb|--extra-options=EXTRA.yaml| can be used to access and update this dictionary. Here, \verb|EXTRA.yaml| is a file in the YAML\footnote{See \url{http://www.yaml.org}}  format containing pairs of keywords and values. For details to the \verb|DarkOptions|  dictionary we refer to the documentation of \DarkAges.

\section{Exotic energy injection within \ExoCLASS~and call to \DarkAges}
\label{sec:ExoClass}
\begin{figure}[htb]
\centering
\includegraphics[width=.95\textwidth]{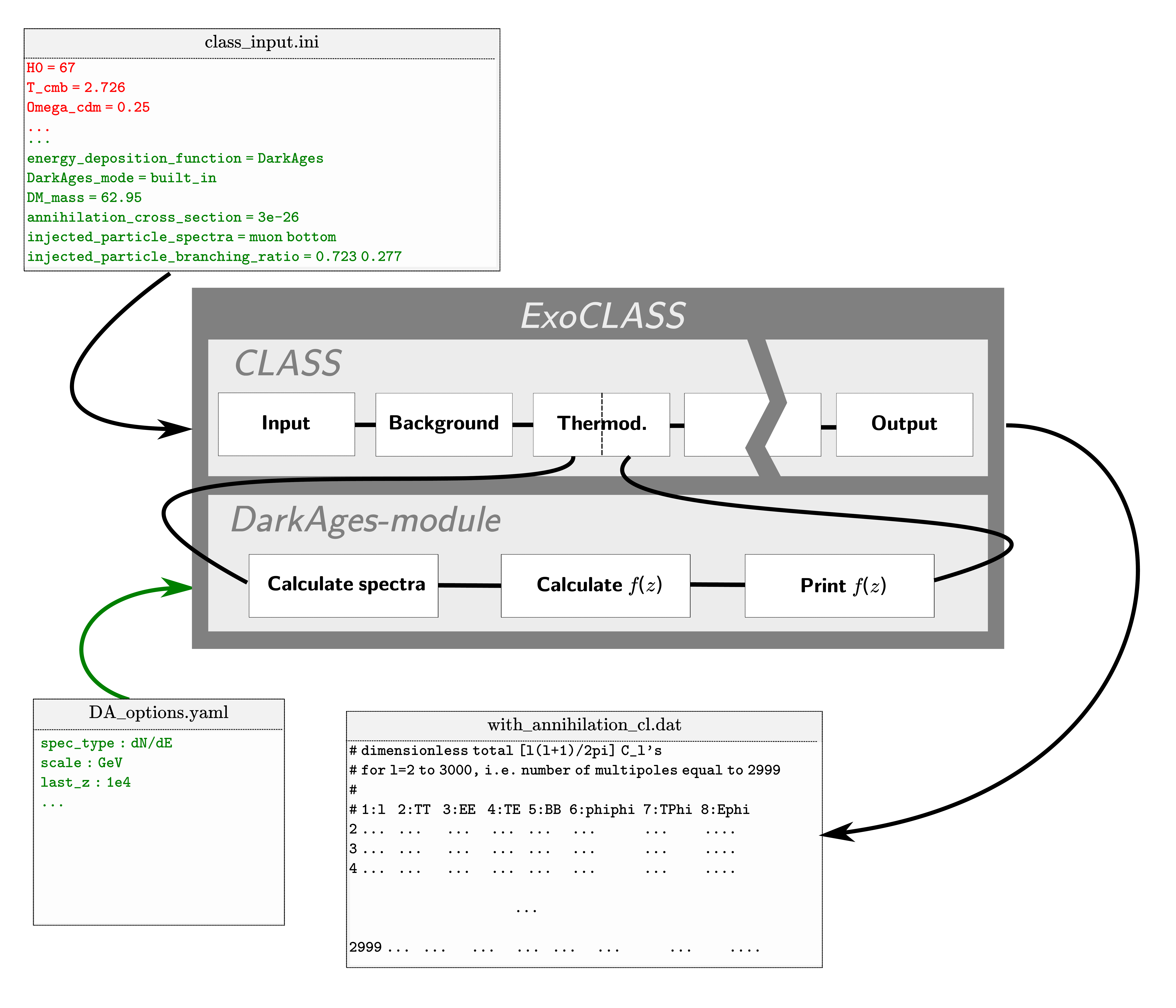}
\caption{\label{fig:DarkAges-flow}Program flow of the \ExoCLASS{}  package ($\mathrel{\widehat{=}}$ \CLASS~interfaced with the \DarkAges  module). In the initialization file the cosmological parameters (in red) and the dark matter model parameters (in green) are given. \CLASS~calls the custom command of the \DarkAges  package within \texttt{thermodynamics\_init()} and fills a table with the output of the computation. After that \CLASS~resumes as usual. In this example, the user is computing the CMB power spectra assuming that the DM annihilates with a cross-section of $3\times10^{-26}$cm$^3$/s into muon and bottom wih branching ratio 0.723 and 0.277 respectively.}
\end{figure}

Within \ExoCLASS, there are several ways to compute the impact of electromagnetic energy injection, typically leading to different level of accuracy/runtime. In order to call a specific energy injection history and treatment of the energy deposition, one needs to specify the following options in the standard input file.\\
\\
{\em Key parameters describing the injection history}: The type of injection history (as described in section \ref{sec:2}) will be set by key parameters of each history. Those are: 
\begin{itemize}
\item For annihilation, the DM mass and cross-section respectively specified with \texttt{DM\_mass} (in GeV) and \texttt{annihilation\_cross\_section} (in cm$^3$/s). Alternatively, the user can choose to pass the usual annihilation parameter $p_{\rm ann}\equiv\sigmav/\mDM$ [m$^3$/s/kg],  called \texttt{annihilation} within \CLASS.  Moreover, when calling the \DarkAges~module (its usage within \ExoCLASS{} will be described below), one can specify the injection spectra and corresponding branching ratio thanks to the options \texttt{injected\_particle\_spectra} and \texttt{injected\_particle\_branching\_ratio} respectively. The usage is the same as the \texttt{spectrum} and \texttt{branching} options described in sec.~\ref{sec:DarkAges} when \DarkAges{} is used as a stand-alone routine. \\
\item For annihilation in halos, one can additionnally pass the amplitude of the halo with the option \texttt{annihilation\_f\_halo} and the redshift at which halos start to form with \texttt{annihilation\_z\_halo}. Their exact definition can be found in Ref.~\cite{Poulin2015}, appendix C. \\
\item For decay, the DM mass, lifetime and fraction (defined with respect to the total DM). Those are specified respectively by  \texttt{DM\_mass} (in GeV), \texttt{tau\_dcdm} (in s) and \texttt{decay\_fraction}. Alike the annihilation case, one can additionnally pass the injected particle spectra and branching ratio when calling \DarkAges. \\
\item For PBH evaporation, the PBH mass (in g) is specified via \texttt{PBH\_evaporating\_mass} and the PBH fraction of the total DM via \texttt{PBH\_fraction}.\\
\item For PBH accretion, the PBH mass (in $M_\odot$) is specified via \texttt{PBH\_accreting\_mass}, the PBH fraction of the total DM via \texttt{PBH\_fraction}. Additionnaly, on needs to specify the accretion recipe thanks to the parameter \texttt{PBH\_accretion\_recipe}, which can be one of \texttt{spherical\_accretion} or \texttt{disk\_accretion}.

\end{itemize}
{\em Parameters setting the treatment of the energy deposition}: In \ExoCLASS, it is possible to pass several parameters adjusting the treatment of the energy deposition. \\
\begin{itemize}
\item \texttt{on the spot}: First and foremost, just like in \CLASS, one can decide to work in the ``on-the-spot'' approximation by setting the parameter \texttt{on the spot = yes}.  As recalled in section \ref{sec:2}, this approximation essentially consists in assuming that the deposition happens at the injection redshift. It is known to be a good approximation in the context of DM annihilation or long-lived decay as long as the injected energy is weighted by a factor $f_{\rm eff}$ adequate for the model under study. There are several ways to specify this parameter. One can first set a constant parameter \texttt{f\_eff} within the input file. However, given the degeneracy of this parameter with other parameters describing the energy injection, one can simply give the parameters relevant to  the energy injection (\texttt{annihilation\_cross\_section}, \texttt{decay\_fraction}, \texttt{PBH\_fraction}) {\em rescaled} by the value of \texttt{f\_eff}. In that case \texttt{f\_eff} can be set to one, or ignored since it is the default value. This is for instance useful in the context of a MCMC runs with \texttt{MontePython}. Finally, one can follow the approach of Refs.~\cite{Slatyer09} later called ``3 keV prescription'' in Ref.~\cite{Slatyer15-2}, which amounts in factorizing the energy deposition function into a redshift-dependent function (describing the high-energy cooling) and another free-electron-fraction dependent one (describing the repartition in each channel). To do so, one can specify the path to a file containing the redshift-dependent $f_{\rm eff}$ in the form
\begin{center}
  \begin{tabular}{|cc|}
  \toprule
  \multicolumn{2}{|c|}{number of lines} \\
    \multicolumn{2}{|l|}{} \\
  $\#z$ &$f_{\rm eff}$ \\
  \ldots & \ldots \\
  \bottomrule 
  \end{tabular}
  \end{center}  
thanks to the command \texttt{energy injection f\_eff file=...}. This approach has for instance been used in Refs.~\cite{Diamanti:2013bia,Lopez-Honorez:2013lcm,Poulin2015,Poulin2016}.
\item \verb|energy_repartition_treatment|: When working in the effective on the spot picture, it is necessary to specify {\em energy repartition functions} often noted $\chi_c(x_e)$. In \CLASS~we have implemented an old prescription advocated by Shull, Van Steenberg, Chen and Kamionkowski (SSCK) \cite{Shull:1982zz,Chen2003}  in which the energy repartitions functions are $\chi_i(x_e)=\chi_\alpha(x_e) = (1-x_e)/3$ and $\chi_h(x_e) = 1 - (\chi_\alpha(x_e)+\chi_i(x_e)) = (1+2x_e)/3$. In this prescription there is no ``lost'' photons $\chi_{<10.2~{\rm eV}} = 0$ and no differentiation between hydrogen and helium reionisation. There is also the possibility to use a more accurate description as computed by Galli, Slatyer, Valdes and Iocco (GSVI) \cite{Galli13}. The user can also specified the path to a file containing a table of the form
\begin{center}
  \begin{tabular}{|cccccc|}
  \toprule
  \multicolumn{6}{|l|}{number of lines} \\
    \multicolumn{6}{|l|}{} \\
  $\#x_e$ & 	$\chi_{\rm heat}$ &	$\chi_{{\rm ly}\alpha}$	&$\chi_{\rm ion-H}$ & 	$\chi_{\rm ion-He}$ & $\chi_{<10.2~{\rm eV}}$ \\
  \ldots & \ldots & \ldots & \ldots & \ldots & \ldots \\
  \bottomrule 
  \end{tabular}
  \end{center}  
 These prescriptions can be chosen by setting \verb|energy_repartition_treatment =| \texttt{SSCK}, \texttt{GSVI} or \texttt{from\_file}.  The path to the file is specified within the input file via the parameter \verb|energy repartition coefficient file=...| .

When working beyond the on-the-spot approximation, it can be useful to by-pass the energy repartition function, since those can be directly included in the energy repartition functions $f_c(z)$. This is done by setting \verb|energy_repartition_treatment =| \verb|no_factorization|, in which case the $\chi$'s functions are effectively set to one within the code.
\item \verb|energy_deposition_function:| This option can be used when working beyond the on-the-spot treatment, i.e. when one sets the option \texttt{on the spot = no} in the input file. The argument can be one of the following
\begin{itemize}
\item \texttt{Analytical\_approximation}: One can first use an analytical approximation developed in Refs.~\cite{Cirelli09,Hooper09,Natarajan09,Giesen} and corrected in Refs.~\cite{Poulin2015,Ali-Haimoud:2016mbv}. A comparaison of this analytic approximation with the full treatment in the context of annihilation in halos is done in appendix B of Ref.~\cite{Poulin2015}.
\item \verb|from_file|: The user might find useful to give $f_c(z)$ functions from an existing file. Its location needs to be given with \verb|energy deposition function file=...|. The file needs to be in the form
 \begin{center}
  \begin{tabular}{|cccccc|}
  \toprule
  \multicolumn{6}{|l|}{number of lines} \\
    \multicolumn{6}{|l|}{} \\
  $\#z_{\rm dep}$ & 	$f_{\rm heat}$ &	$f_{{\rm ly}\alpha}$	&$f_{\rm ion-H}$ & 	$f_{\rm ion-He}$ & $f_{<10.2~{\rm eV}}$ \\
  \ldots & \ldots & \ldots & \ldots & \ldots & \ldots \\
  \bottomrule 
  \end{tabular}
  \end{center}  
\item \verb|DarkAges|: Last but not least, it is possible to call the \DarkAges~package by setting \verb|energy_deposition_function=DarkAges| in the input file.  In that case, \ExoCLASS{} automatically sets the mandatory parameters \verb|on the spot = no| and \verb|energy_repartition_treatment = no_factorization|.
\end{itemize}

\end{itemize}
{\em Calling the {\em \DarkAges}~module}: Apart from its use as a stand-alone package to compute the $f_c(z)$ functions, the \DarkAges-module is designed to be used in conjuction with \CLASS. It is called by setting \verb|energy_deposition|
\verb|_function=DarkAges| in the input file.
In figure \ref{fig:DarkAges-flow} we show the program-flow of \ExoCLASS: all relevant parameters, i.e. the cosmological parameters used by \CLASS, as well as the input parameters relevant for the correct input of the \DarkAges-comandline script, are passed as input parameters of \CLASS{} in a standard \verb|.ini|-file. The call to the module is done in the \texttt{thermodynamics.c} file. \DarkAges~then computes and return the $f_c(z)$ functions to \ExoCLASS{} which are used only within this module. If needed, the user can print the output of the computation in a file with the option \verb|print_energy_deposition_function = yes|.

 The \DarkAges-module can be used in several modes, which are specified via the option \texttt{DarkAges\_mode = MODE}. Those were described in previous subsection and are called within \ExoCLASS{} with the following keywords
\begin{itemize}
 
 \item \verb|built_in|: In this mode, \ExoCLASS~sets up an automatic call to the \DarkAges-script given the scenario of energy injection the user is interested in. From the key parameters of the injection history that are given to \ExoCLASS, the code automatically detects what history the user is computing.
 \item \verb|user_command|: If the call to the \DarkAges-script cannot be performed with one of the \texttt{built-in} methods, this setting allows the user to have full control on the python commandcalled by \ExoCLASS{} to get the $f_c(z)$-table. The command is given via the option \verb|ext_fz_command=...| and there is the possibility to use up to five additional floating point parameters \verb|ext_fz_par1|, \ldots, \verb|ext_par_fz5| which are meant to be used as varying parameters when using \ExoCLASS{} with \texttt{MontePython} (those can be for instance some relevant couplings). The full command which is executed by \CLASS{} is then set up by the content of \verb|ext_fz_command| and the five additional values \verb|ext_fz_par|$i,\,i =1,2,3,4,5$ are appended to it separated by blanks. 
\end{itemize}

\begin{figure}
\centering
\includegraphics[scale=0.4]{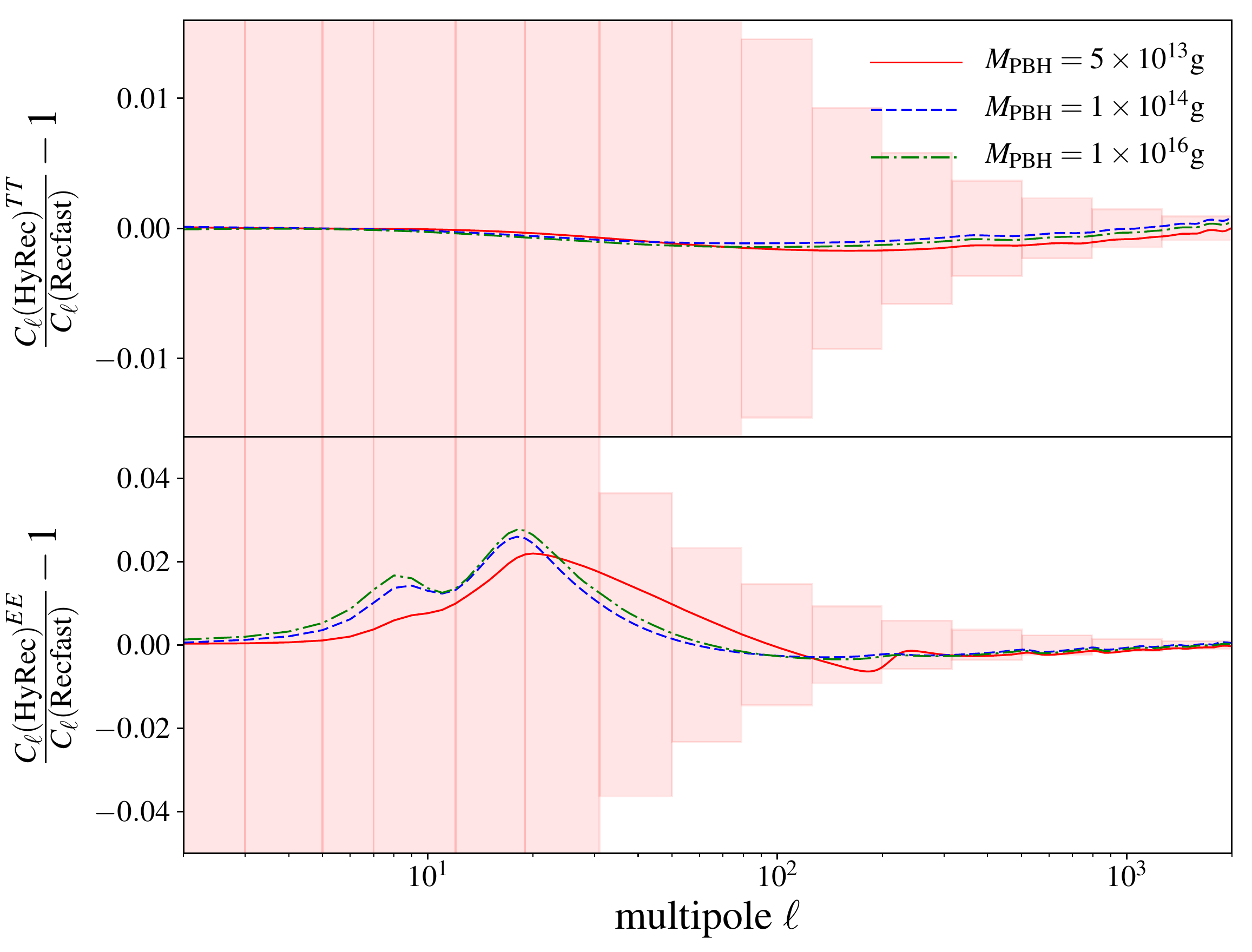}
\caption{Residuals of the CMB TT and EE power spectra computed using \texttt{Recfast} and \texttt{HyRec} or three different PBH masses. The abundance set by the constraint at 95\%C.L. obtained in this work.}
\label{fig:Hyrec_vs_recfast}
\end{figure}

Thanks to clever preparation and interpolation routines, it is perfectly possible to perform intense MCMC runs using \ExoCLASS{} and \texttt{MontePython}.  \ExoCLASS~is especially fast in conjuction with the fuged version of \texttt{Recfast} implemented in \CLASS, which has been shown to be accurate enough to describe the recombination history from detailed comparaison with \texttt{CosmoRec} and \texttt{HyRec}.  This is illustrated on figure \ref{fig:Hyrec_vs_recfast} in the context of PBH evaporation, where we plot the residuals of the CMB TT and EE power spectra computed using \texttt{Recfast} and \texttt{HyRec} for three different PBH masses and abundance set by the constraint at 95\%C.L. obtained in this work. The difference is reassuringly well below cosmic variance.

\bibliographystyle{ieeetr}

\bibliography{references}
\end{document}